\documentclass[reprint,onecolumn,aps,pra,notitlepage,superscriptaddress]{revtex4-1} 
\usepackage{amsmath, amssymb, amsfonts}
\usepackage{graphicx}
\usepackage{color}

\def\rmd{\mathrm{d}}
\def\rmi{\mathrm{i}}

\def\rms{\mathrm{s}}

\def\bfx{\textbf{\em x}}
\def\bfd{\textbf{\em d}}
\def\bfp{\textbf{\em p}}
\def\bfk{\textbf{\em k}}
\def\bfE{\textbf{\em E}}
\def\bfA{\textbf{\em A}}
\def\bfeps{\bfgreek{\epsilon}_\mathrm{HHG}}

\def\tr{{t_\mathrm{r}}}
\def\ti{{t_\mathrm{i}}}
\def\Ip{{I_\mathrm{p}}}
\def\CEP{\varphi_\mathrm{CEP}}

\def\ki{[\bfp_\rms+\bfA(\ti)]}
\def\kr{[\bfp_\rms+\bfA(\tr)]}

\def\bfgreek#1{\mbox{\boldmath{$#1$}}}
 
\def\abs2#1{\left|#1\right|^2}

\newcommand{\LOA}{Laboratoire d'Optique Applique\'e, ENSTA-Paristech, Ecole Polytechnique, CNRS, 91761 Palaiseau Cedex, France}
\newcommand{\PIVUT}{Photonics Institute, Vienna  University of Technology, Gu\ss hausstra\ss e 27/387, 1040 Vienna, Austria}
\newcommand{\IC}{The Blackett Laboratory, Imperial College London, South Kensington Campus, London SW7 2AZ, United Kingdom}

\begin{document}

\title{Enhanced multi-colour gating for the generation of high-power isolated attosecond pulses}

\author{S. Haessler}\email{stefan.haessler@ensta-paristech.fr}\affiliation{\LOA}\affiliation{\PIVUT}
\author{T.  Bal\v{c}iunas}\affiliation{\PIVUT}
\author{G. Fan}\affiliation{\PIVUT}
\author{L. E. Chipperfield}\affiliation{\IC}
\author{A. Baltu\v{s}ka}\affiliation{\PIVUT}

\date{\today}

\begin{abstract}
Isolated attosecond pulses (IAP) generated by high-order harmonic generation are valuable tools that enable dynamics to be studied on the attosecond time scale. The applicability of these IAP would be widened drastically by increasing their energy. Here we analyze the potential of using multi-colour driving pulses for temporally gating the attosecond pulse generation process. We devise how this approach can enable the generation of IAP with the available high-energy kHz-repetition-rate Ytterbium-based laser amplifiers (delivering 180-fs, 1030-nm pulses). We show theoretically that this requires a three-colour field composed of the fundamental and its second harmonic as well as a lower-frequency auxiliary component. We present pulse characterization measurements of such auxiliary pulses generated directly by white-light seeded OPA with the required significantly shorter pulse duration than the that of the fundamental. This, combined with our recent experimental results on three-colour waveform synthesis [Phys. Rev. X 4, 021028 (2014)], proves that the theoretically considered multi-colour drivers for IAP generation can be realized with existing high-power laser technology. The high-energy driver pulses, combined with the strongly enhanced single-atom-level conversion efficiency we observe in our calculations, thus make multi-colour drivers prime candidates for the development of unprecedented high-energy IAP sources in the near future.
\end{abstract}

\maketitle

\section{Introduction}

More than a decade ago, the first observations of isolated attosecond light pulses~\cite{Hentschel2001Attosecond} (IAP) or attosecond pulse trains (APT)~\cite{Paul2001Observation,Hentschel2001Attosecond}  launched the field of ``attosecond physics'', for which these optical tools became the technological base. The high-order harmonic generation (HHG) process in atoms, at the origin of these attosecond light pulses, can be described by a semi-classical three-step model~\cite{Schafer1993,Corkum1993Plasma}, where an electron is first released by tunnel ionization, then accelerated by the strong laser field and finally driven back to the core. The recombination leads to the emission of an ultrashort burst of XUV light. In a multi-cycle laser pulse, this process is repeated coherently every half cycle, leading to the emission of an APT. 

In order to isolate a single attosecond pulse and thus greatly facilitate the implementation of traditional pump-probe schemes for time-resolving processes on the sub-femtosecond time scale, it is necessary to temporally gate the HHG process and limit it to a single laser half-cycle~\cite{Calegari2012gatingreview,Chini2014review}. For example, using carrier-envelope-phase (CEP) controlled near-single-cycle driver pulses ($\lesssim4\:$fs for an 800-nm carrier wavelength), the highest XUV photon energies are only produced by a single driver half-cycle. Spectrally selecting this cutoff region then selects an IAP~\cite{Hentschel2001Attosecond,Goulielmakis2008SingleCycle}. Slightly longer driving pulse envelopes and a wider selectable XUV spectral range can be used with polarization-shaping, such that the driver pulses have elliptically polarized wings and a linearly polarized center. This allows recollision of the laser-driven electron trajectories only in a single-half cycle in the pulse center~\cite{Tcherbakoff2003,Sola2006Controlling}. Using driver pulses with very high intensity leads to a rapidly rising degree of medium ionization already during the rising edge of the pulse. With few-cycle envelope duration, this may leave a narrow time-window of about one driver half-cycle during which the efficient XUV generation remains possible (through a combination of efficient tunnel ionization launching electron trajectories, transient phase-matching with a tolerable free-electron density,and depletion of the ground state)~\cite{Pfeifer2007,Ferrari2010highenergy}. Finally, a sufficiently fast wavefront rotation of the driver pulses leads to the individual attosecond pulses of an APT to be emitted at varying angles, such that an IAP may be selected by a spatial mask in the far field~\cite{Quere2014LighthouseApplications,Heyl2014noncollinear}.

All these techniques require extremely short $\lesssim2$-cycle driver pulse durations. Due to gain-narrowing, these cannot be achieved directly from laser amplifiers, so an additional post-compression stage after the amplifier is required, where the laser spectrum broadens by self-phase modulation during the propagation through a gas-filled hollow-core fiber and is subsequently compressed with chirped mirrors~\cite{Nisoli1996hcfiber,Andriuk2011hollow}. This typically limits the driver pulse energy to a few mJ~\cite{Boehle2014sallenoit4fs} due to the onset of ionisation. Even with the best optimization effort, the attainable IAP energy is thus limited to the nJ- level simply through the limited driver pulse energy.

While very advanced OPCPA schemes with careful phase control do permit the generation of a sub-3-cycle pulses with very high pulse energy~\cite{Herrmann2009} or average power~\cite{Rothhardt2012fewcycleOPCPA}, this complex technology remains rather inefficient. 

The resulting low power / pulse energy of current IAP sources hampers a much wider applicability of current attosecond science methods to lower sample densities (like molecules, clusters or nanoparticles that have to be evaporated or ablated to be brought into the gas phase, or excited species that only exist as a small fraction of the target medium) or to processes with low cross-sections. It also makes it very hard to reach a non-linear interaction regime for the XUV attosecond pulses~\cite{Tzallas2003,Takahashi2013gigawattatto}. 

The gating technique known as multi-colour gating has lead to a breakthrough by enabling IAP generation with $\approx10$-cycle drivers. It is based on the coherent mixing of two or more laser pulses with different carrier wavelengths (and usually parallel linear polarization directions). Already in 2006, it was demonstrated that the broken symmetry of a two-colour field composed of a fundamental with its second harmonic doubles the spacing of attosecond pulses from one half-cycle to one full laser cycle (2.7~fs in the case of an 800-nm fundamental), which correspondingly allows generating IAP with twice as long (i.e. $\approx4$-cycle) driver pulses~\cite{Pfeifer2006heterodyne,Oishi2006sub10twocolor}. The usable driver duration was further increased to $\approx10$-cycles by either adding polarization shaping to this scheme to perform ``double optical gating''~\cite{Feng2009generation}, or by detuning the second harmonic~\cite{Merdji2007}, which further increases the periodicity of the two-colour field.

While such ``incommensurate'' two-colour combinations turned out to be barely realizable by optical harmonic generation with its very limited tunability, optical parametric amplification (OPA), creating a frequency down-converted additional colour, lends itself ideally to this task. The price to pay is the fact that the phase-delay between the fundamental and the auxiliary colour component is not passively stable, as in optical harmonic generation. Using a white-light-seeded OPA~\cite{Baltuska2002} together with an actively CEP-locked fundamental (pump) laser does however yield a stable phase delay between the pump and the signal / idler wave and thus stable multi-colour waveforms~\cite{Haessler2014perfectwave}. Without CEP-locking, the phase delay fluctuates and the gating efficiency thus varies shot-to-shot~\cite{Vozzi2009,Siegel2010smurf}. Conditions can however be found, notably including sufficiently short pulse envelopes, which yield an IAP for a very wide CEP-range~\cite{Takahashi2010,Takahashi2013gigawattatto}, thus relaxing the CEP-locking requirement, albeit for the price of strong shot-to-shot fluctuation of the IAP intensity. The latter method yielded record-high $\mu$J-level energy per IAP, constituting a 100-times enhancement compared to previous achievments~\cite{Ferrari2010highenergy} resulting mainly from the fact that the longer driver pulses can be taken directly from laser amplifiers or OPA.
Note that multi-colour gating is much more energy-efficient than polarization- or ionization-based gating techniques. The latter ``waste'' a lot of field-energy in parts of the pulse where HHG is shut off by elliptical polarization or excessive ionization, whereas the multi-colour beating automatically concentrates the available field energy in a few very strong field oscillations which then efficiently drive the desired HHG process. There are even indications that undesired ionization might be reduced~\cite{Takahashi2010}, which would allow reaching higher saturation intensities and thus higher HHG efficiency and cutoff. 

Here, we are going expand on the temporal gating aspect of our recent work on optimizing HHG at the single-atom-level through shaped multi-colour driver waveforms~\cite{Haessler2014perfectwave}. We derive a clear understanding from a simple time-domain picture and then demonstrate, using single-atom-level numerical simulations with the Lewenstein model of HHG~\cite{Lewenstein1994Theory}, how the multi-colour gating technique combined with (rough) CEP-control so that stable multi-colour waveforms are realized enables the generation of IAP based on 180-fs long pulses typically supplied by Yb-based laser amplifier technology. The extremely efficient temporal gating is accompanied by strongly enhanced IAP intensities compared to a single-colour driver, achieved through advantageous shaping of the central driving waveform cycle as experimentally demonstrated in our recent work~\cite{Haessler2014perfectwave}. Our results suggest that high-energy / high-flux sources of isolated attosecond pulses based on Ytterbium-laser technology without a need for pulse-post-compression are within immediate reach. Finally, we discuss the experimental implementation. 

\section{Theoretical analysis}

\subsection{Temporal gating by two-colour ``few-cycle-pulse trains'' \label{sec:fewcycletrains}}

A formal dynamical symmetry analysis~\cite{Fleischer2006bichromatic} shows that incommensurate two-colour combintations, i.e. of fields with frequencies  $\omega_0$ and $\omega_1=\eta\omega_0$  with $0.5<\eta<1$, leads in the spectral domain to a densification of high-harmonic peaks---independently of the colour-components' phase delay. The derived spectral selection rules have been experimentally verified~\cite{Bandulet2010} and the potential for temporal gating of HHG has been underlined.

While this potential has already been examined in a number of (parametric) numerical studies published during the last years~\cite{Kim2008optimization,Zou2010chirpedtwocolor,Zhang2014twocolor}, we will here, in contrast, base our reasoning on an intuitive time-domain picture, inspired by a brief mention in  ref.~\cite{Fleischer2006bichromatic}. We find this picture to give a very accessible understanding of the temporal gating effect and its dependence of the colour components' phase delay, which enables us to devise and understand very easily the optimal colour combinations for generating an IAP over a broad CEP-range.

Coherently combining two laser waves with equal strength and frequencies $\omega_0$ and $\omega_1$ leads to a beating which can be understood by considering
\begin{equation}
	\cos(\omega_0 t + \varphi_0 ) + \cos(\omega_1 t + \varphi_1 ) = 2 \cos\left( \frac{\omega_0 - \omega_1}{2} t + \frac{\varphi_0 - \varphi_1}{2}\right)  
										\:  \cos\left( \frac{\omega_0 + \omega_1}{2} t + \frac{\varphi_0+\varphi_1}{2}\right), 
\label{eq:factorization}
\end{equation}
i.e. the factorization into a slowly oscillating ``envelope'' term and a faster oscillating ``carrier'' term. We are particularly interested in frequency ratios $0.5<\eta<0.9$ since these produce waveforms that can be considered as a train of few-cycle pulses with carrier frequency $(\omega_0 + \omega_1)/2$ and CEP $(\varphi_0+\varphi_1)/2$, spaced by a period $2\pi/(\omega_0 - \omega_1)$.

\begin{figure}
	\includegraphics[width=1\textwidth]{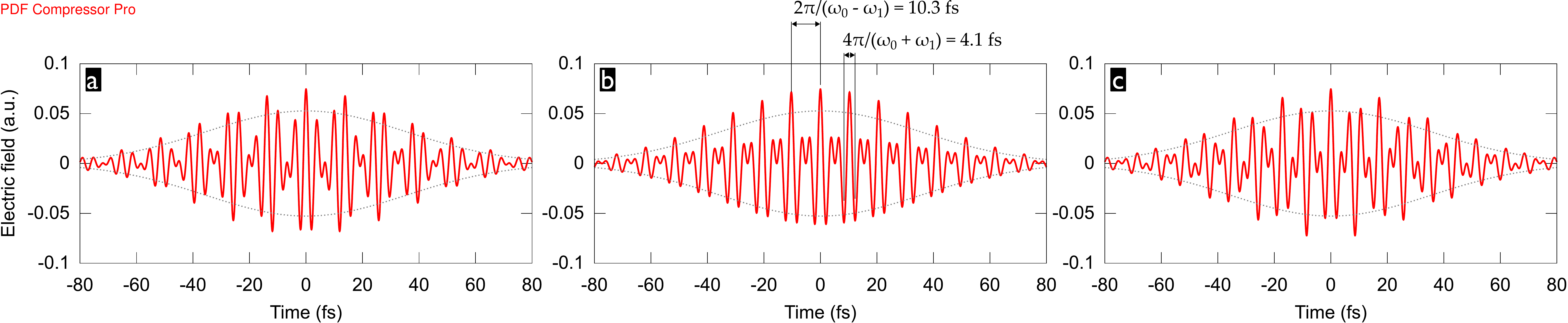}
	\caption{Electric field (red line) of two-colour pulses composed from a 1030-nm fundamental and a 1400-nm ($N=6.6$) (a), 1545-nm ($N=5$) (b), or 1700-nm ($N=4.1$) (c) auxiliary component, each with a 60-fs FWHM intensity envelope and a peak intensity of $0.5\times10^{14}\:\mathrm{W\,cm^{-2}}$. $\CEP=0$.  For comparison, the gray dashed lines show the field-envelope of a 60-fs single colour pulse with $1\times10^{14}\:\mathrm{W\,cm^{-2}}$ peak intensity.}
	\label{fig:factorization}
\end{figure}

An important parameter is the ratio 
\begin{equation}
 N = \frac{\omega_0 + \omega_1 }{\omega_0 - \omega_1},
\end{equation}
i.e. twice the number of carrier-cycles per few-cycle sub-pulse. For integer $N$, the few cycle pulses repeat with constant CEP when N is odd  (i.e. the waveform is strictly $2\pi/(\omega_0 - \omega_1)$-periodic, see figure \ref{fig:factorization}b), or with a $\pi$ CEP flip between successive pulses when $N$ is even (see figure \ref{fig:factorization}c). For non-integer $N$, there is a corresponding fraction-of-$\pi$ CEP slip between successive few-cycle pulses (see figure \ref{fig:factorization}a). For targeting a particular ratio $N$, the relation
\[
\eta=(N-1)/(N+1)
\]
 will be useful.

Since we consider the $\omega_1$-auxiliary-component to be generated from a white-light-seeded OPA pumped by the $\omega_0$-fundamental-component, we set $\varphi_0=\varphi_1=\CEP$ so that the phase delay between the two colour components, $\tau_\mathrm{\phi} = \CEP({\omega_1}^{-1} - {\omega_0}^{-1})$, is CEP-dependent. The ``envelope'' term in equation (\ref{eq:factorization}) is then always maximal at the pulse center, $t=0$, and the ``carrier'' term has a CEP of $\CEP$. This means that the phase delay variation caused by a drift of $\CEP$ can also be thought of as a CEP-drift of the few-cycle sub-pulses. 

%
Figure \ref{fig:factorization}b shows an illustrative example of two-colour pulses composed of 1030~nm and 1545~nm, which corresponds to the particular case of odd $N=5$. We obtain a train of near-single-cycle sub-pulses with 1236~nm carrier wavelength (corresponding to a 4.1-fs oscillation period) spaced by a 10.3-fs period. Due to the odd $N$, the sub-pulses all have the same CEP, $\CEP$. Even with the rather long 60-fs pulse envelopes, the large sub-pulse spacing leads to a significant variation of the peak field strength from one sub-pulse to the next, which underlines the potential for efficient temporal gating. 

%
Detuning the auxiliary wavelength has a number of coupled consequences. Bringing $\omega_1$ closer to the fundamental $\omega_0$ (see figure \ref{fig:factorization}a) increases the spacing of the sub-pulses (e.g. to 13~fs for a 1400-nm auxiliary wavelength), as well as their effective duration, while decreasing their carrier wavelength  (e.g. to 1187~nm for a 1400-nm auxiliary wavelength)---the sub-pulses thus contain more cycles. Bringing $\omega_1$ farther away from $\omega_0$ has the opposite effect (see figure \ref{fig:factorization}c). For temporal gating this means that there is an optimum auxiliary wavelength to be found where \textit{(i)} the sub-pulses are neither too close, meaning that adjacent ones would contribute to the near-cutoff photon energy range, thus forming an APT, nor \textit{(ii)} where the sub-pulses contain too many cycles of the carrier wave, meaning that within the central sub-pulse, more than one attosecond pulses in the near-cutoff energy range would be generated. Point  \textit{(i)} can be mitigated by an $\omega_1$-value correponding to a non-integer $N$, so that the adjacent sub-pulses experience a CEP-shift. If the central sub-pulse has an optimal CEP for efficiently generating high photon energies, the adjacant ones would then be sub-optimal and their contribution suppressed  (cp. figure \ref{fig:factorization}a). 

This picture is complementary to the one proposed in \cite{Tosa2012isolated} where a ``phase accordance'' parameter of the two colour components is defined, which takes small values in gating windows corresponding to our sub-pulses. The authors of \cite{Tosa2012isolated} did however not study the experimentally very important CEP-dependence of the gating and generation efficiency.

\subsection{Numerical simulations}

For our numerical simulations with the non-adiabatic quantum path analysis of the Lewenstein model for HHG~\cite{Lewenstein1994Theory,Sansone2004nonadiabatic} (see Methods-section), we start with 60-fs FWHM-intensity envelopes for all colour components before exploring even longer durations. We consider a total peak intensity (i.e. the sum of the peak intensities of the colour components) of $I_\mathrm{tot} = 1\times10^{14}\:\mathrm{W\,cm^{-2}}$. This means that, unless stated otherwise, we always compare driver pulses of the same total pulse energy. While we consider the ionization potential of argon atoms, $\Ip= 15.76\:$eV, note that hydrogenic dipole matrix elements $\bfd [\bfk]$ are used so as to avoid an influence of the particular atomic structure on our conclusions. The driving fields are always linearly polarized and all colour components have parallel polarization directions. The attosecond emission is calculated from the HHG spectra after application of a 15-eV wide spectral filter selecting the region of highest generated photon energies near the cutoff.

\begin{figure}
	\includegraphics[width=0.8\textwidth]{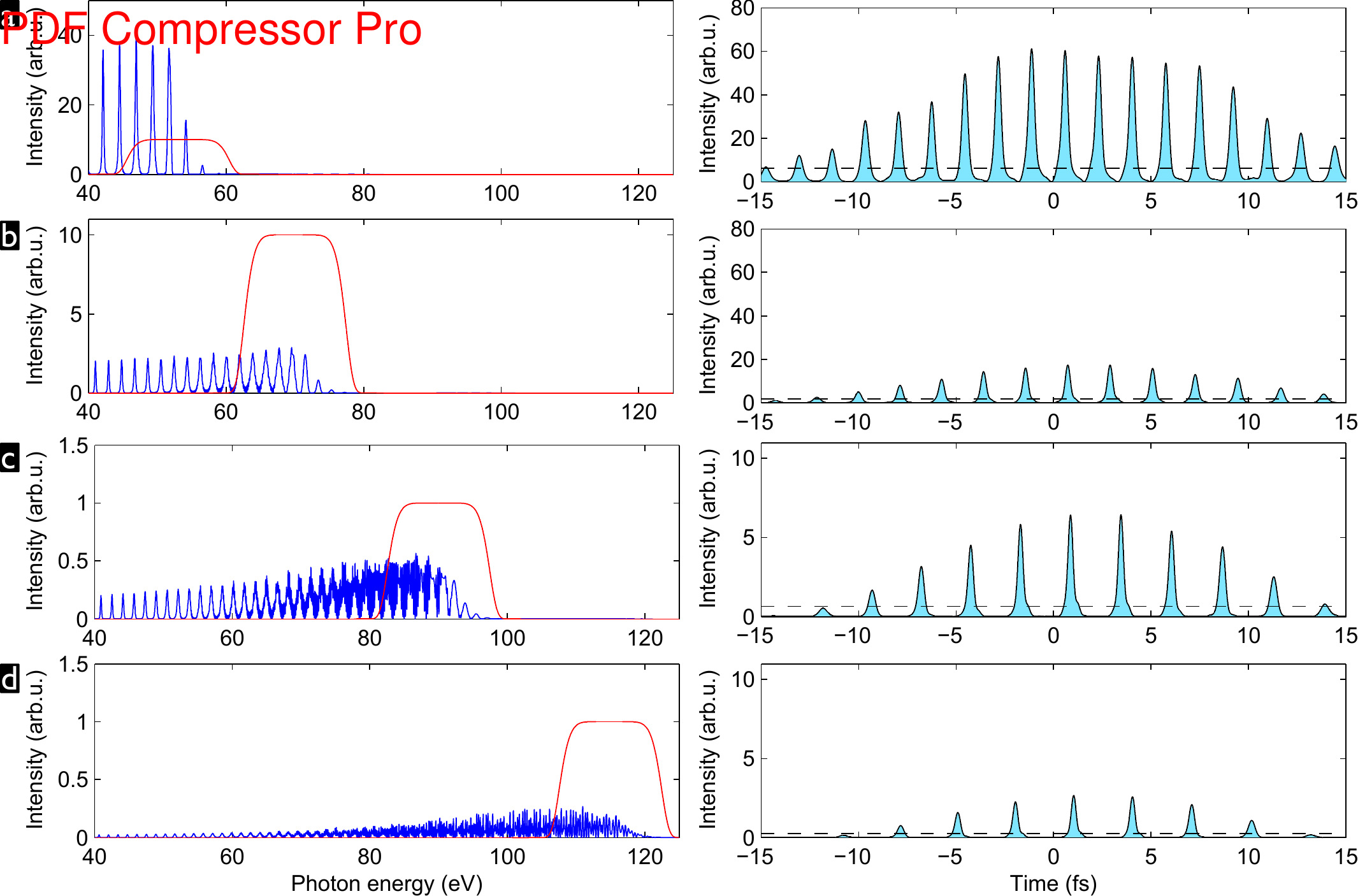}
	\caption{High harmonic spectrum (left) and attosecond pulses (right) for single-colour driver pulses with $\CEP=0$, 60-fs duration and 1030-nm (a), 1300-nm (b), 1545-nm (c), and 1800-nm (d) carrier wavelength. The red line superposed to the spectra shows the 15-eV wide normalized super-Gaussian filter centered at the cutoff frequencies 53~eV (a), 70~eV (b), 90~eV (c), and 115~eV (d), which selects the spectral window corresponding to the shown attosecond pulses. The dashed line marks the 10\%-level of the intensity of the strongest attosecond pulse.}
	\label{fig:singlecolours}
\end{figure}
As a reference, we show in figure \ref{fig:singlecolours} the HHG spectra generated by single-colour drivers with  varying carrier wavelengths, as well as the temporal profile of the attosecond emission. Not surprisingly, with the considered many-cycle pulse duration the emission comes in the shape of an APT, even for the longest wavelength of 1800~nm. The well-known increase of the attainable photon energy cutoff accompanied by a rapid drop of conversion efficiency~\cite{Tate2007,Shiner2009} is also clearly apparent.

\subsubsection{Two-colour gating}

\begin{figure}
	\includegraphics[width=\textwidth]{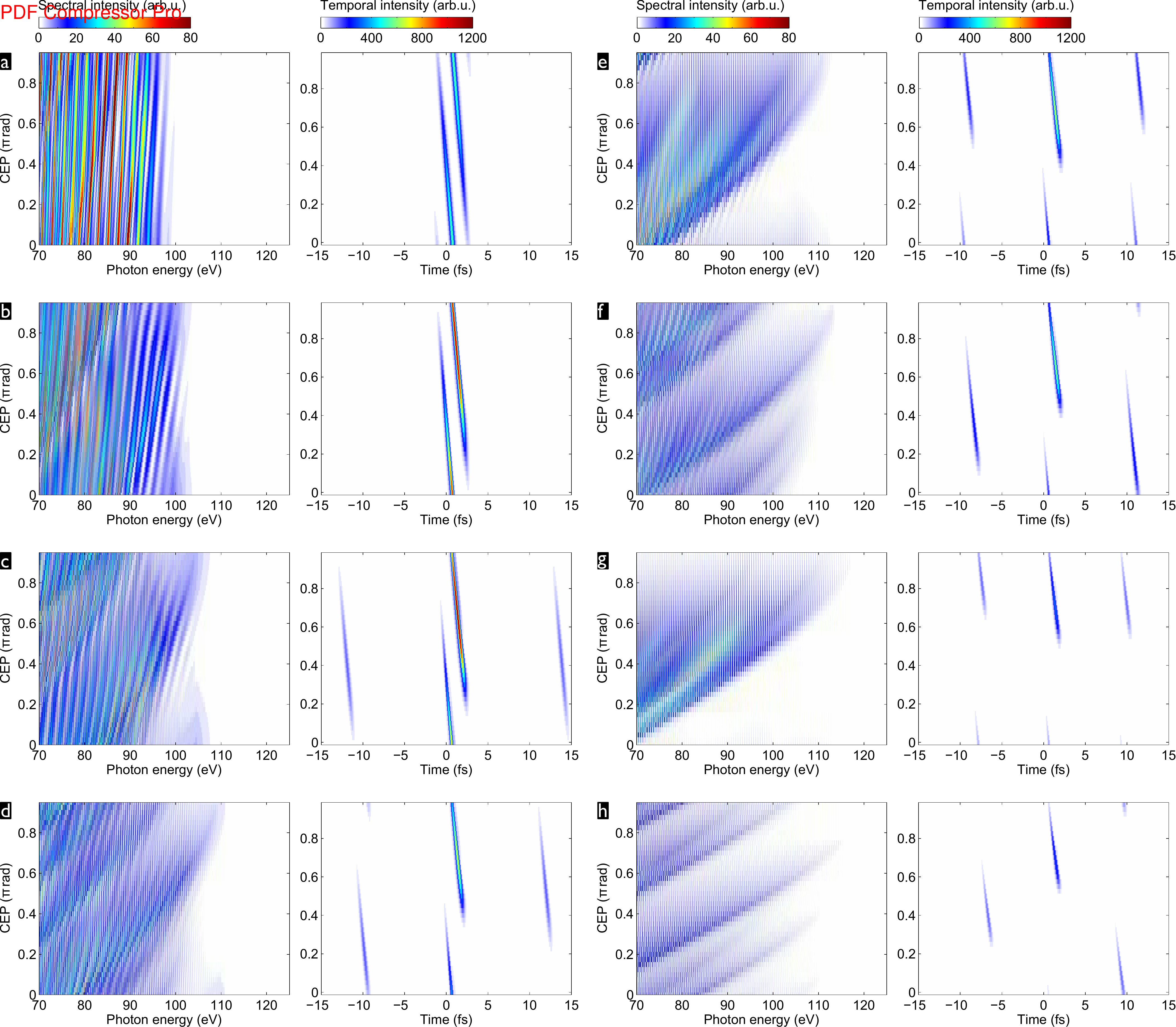}
	\caption{CEP-dependence of the generated high harmonic spectrum (left) and attosecond pulses (right) for the combination of a 1030-nm fundamental with an auxiliary 1200-nm, $N=13.1$ (a), 1300-nm, $N=8.6$ (b), 1400-nm, $N=6.6$ (c), 1500-nm, $N=5.4$ (d), 1545-nm, $N=5$ (e), 1600-nm, $N=4.6$ (f), 1700-nm, $N=4.1$ (g), or 1800-nm, $N=3.7$ (h) pulse, both with a 60-fs duration. The pulses have peak intensities of $I_0 = 0.5\times10^{14}\:\mathrm{W\,cm^{-2}}$ and $I_1 = 0.5\times10^{14}\:\mathrm{W\,cm^{-2}}$, respectively, \emph{i.e. $R=1$}. For the attosecond pulses, the cutoff region has been selected by a 15-eV wide normalized super-Gaussian filter centered at 100~eV (a-c), 105~eV (d-f) or 110~eV (g,h) (cp. figure \ref{fig:selectedCEPsR1}). Outside of the shown time window, the attosecond emission is suppressed to below a $10^{-4}$-fraction of the strongest attosecond pulse's peak intensity.}
	\label{fig:CEPscansR1}
\end{figure}
\begin{figure}
	\includegraphics[width=0.75\textwidth]{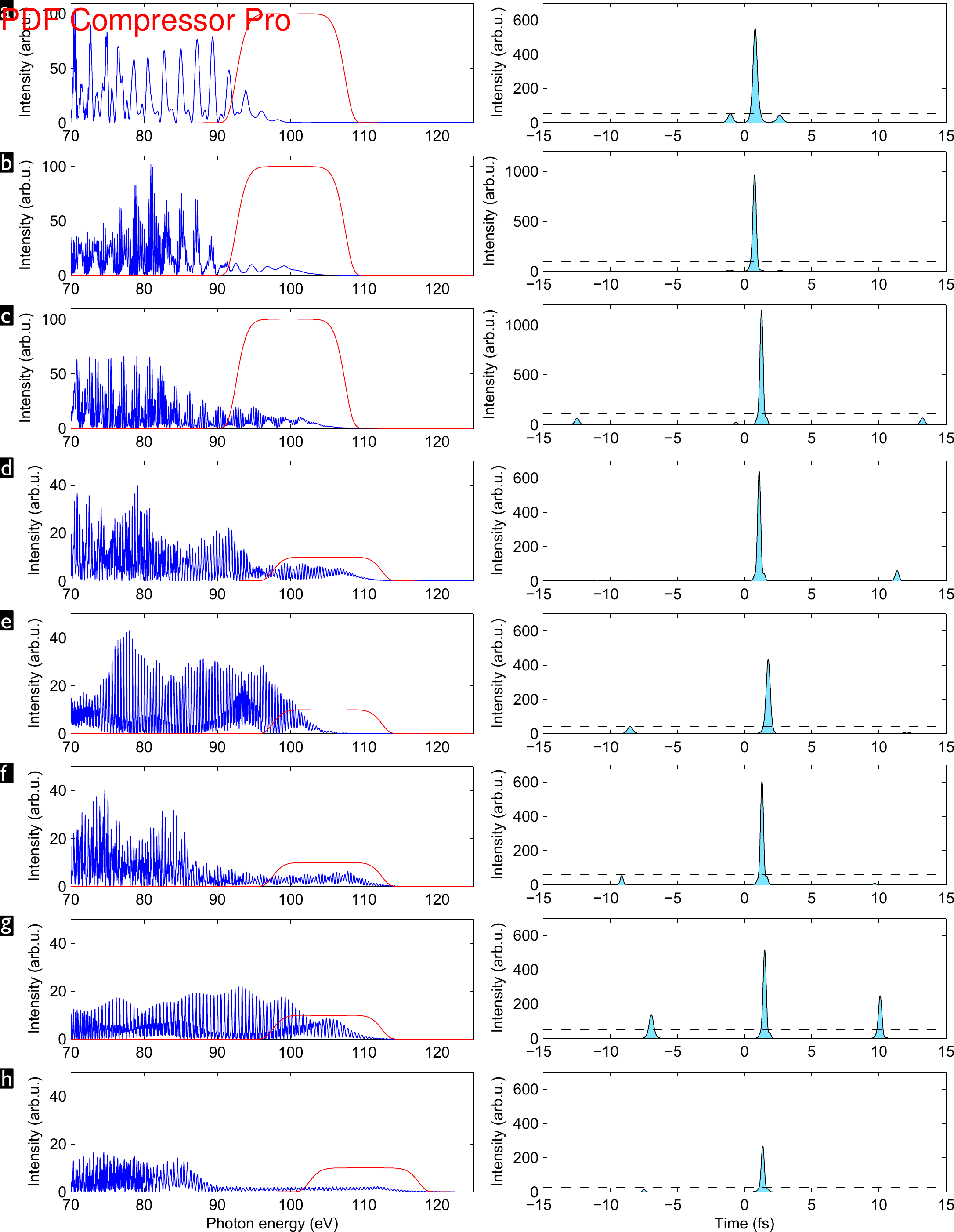}
	\caption{High harmonic spectrum (left) and attosecond pulses (right) for selected optimal phase delays out of the scans in figure \ref{fig:CEPscansR1} ($R=1$): for the auxiliary 1200-nm pulse, $\CEP=0\:$rad (a), for 1300-nm, $\CEP=0\:$rad (b), for 1400-nm, $\CEP=0.7\pi\:$rad (c), for 1500-nm, $\CEP=0.8\pi\:$rad (d), for 1545-nm, $\CEP=0.55\:$rad (e), for 1600-nm, $\CEP=0.7\pi\:$rad (f), for 1700-nm, $\CEP=0.7\pi\:$rad (g), for 1800-nm, $\CEP=0.7\pi\:$rad (h). The red line superposed to the spectra shows the 15-eV wide normalized super-Gaussian filter centered at 100~eV (a-c), 105~eV (d-f) or 110~eV (g,h), which selects the cutoff region for the attosecond pulses. The dashed line marks the 10\%-level of the intensity of the strongest attosecond pulse.}
	\label{fig:selectedCEPsR1}
\end{figure}

We consider two-colour fields of the form
\begin{equation}
	E(t) = f_0(t) \mathrm E_0 \cos(\omega_0 t + \CEP) + f_1(t) E_1 \cos(\omega_1 t + \CEP) ,
	\label{eq:twocolourfield}
\end{equation}
with a peak intensity ratio $R=\abs2{E_1/E_0}$ so that $E_0 = [I_\mathrm{tot}/(R+1)]^{1/2}$ and $E_1 = [I_\mathrm{tot} R/(R+1)]^{1/2}$. The field-envelope functions are set to $ f_0(t)= f_1(t) = \cos^2[ 2\arccos(2^{-1/4})\, t /\tau]$ with the intensity-FWHM duration $\tau=60\:$fs. We consider a fundamental frequency $\omega_0$ corresponding to the emission wavelength of Yb-based lasers of 1030~nm.

Figure \ref{fig:CEPscansR1} shows for $R=1$ the generated high-harmonic spectra and attosecond emission as a function of $\CEP$, i.e. of the phase delay between the two colour components, for a range of auxiliary wavelengths, $\lambda_1$, from 1200~nm to 1800~nm. As the carrier-wavelength of the two-colour waveform increases (with increasing auxiliary wavelength), the achieveable HHG cutoff photon energy increases accompanied by a drop in conversion efficiency, as expected from the well-known wavelength scaling of HHG~\cite{Tate2007,Shiner2009}. As for temporal gating, the trends expected from the discussion of section \ref{sec:fewcycletrains} are clearly observed: \textit{(i)} For the shorter auxiliary wavelengths, the central sub-pulse becomes so long that it generates, for most CEP values, two attosecond pulses separated by a half-period of its carrier wave. As the auxiliary wavelength gets longer, the transition to a single-cycle central sub-pulse is observed, effectively limiting the attosecond pulse generation to a single half-cycle. \textit{(ii)} For the shorter auxiliary wavelengths, the long spacing of the sub-pulses effectively supresses the attosecond emission by the sub-pulses adjacent to the central one. \textit{(iii)} For the auxiliary wavelengths corresponding to (close-to) integer $N$, 1545~nm and 1700~nm, the sub-pulses all have the same CEP and the attosecond emission is correspondingly strongest at the same CEP-values for the central sub-pulse and its adjacent ones. In contrast, the CEP-shift between subsequent sub-pulses for non-integer $N$ is beneficial for temporal gating since it shifts the maxima of the attosecond emission of the adjancent sub-pulses to different CEP values as that of the central sub-pulse (e.g. compare figures \ref{fig:CEPscansR1}d,e,f).

We now select optimal CEP-values from the scans shown in \ref{fig:CEPscansR1} as follows: if there are CEP-values for which the contrast ratio between the intensity $I_\mathrm{max}$ of the strongest attosecond pulse and that of the second-strongest one is $\varrho>10$, then only these are considered, and the CEP with the highest ``fitness'' equal to ${I_\mathrm{max}}^4\varrho$ is selected. This ensures that we enforce a good contrast $\varrho>10$ while giving preference to a attosecond peak intensity. If no CEP-value provides $\varrho>10$, we relax the intensity preference and selected the CEP value with the highest fitness $I_\mathrm{max}\varrho$.

The HHG spectra and attosecond emission so selected is shown in figure \ref{fig:selectedCEPsR1}. With the exception of $\lambda_1=1700\:$nm (due to an unfavorable combination of small driving sub-pulse spacing and near-integer $N$), all considered auxiliary wavelengths permit, within a limited CEP-range, to generate an IAP. Here we define IAP as a single attosecond pulse which dominates over potential pre- or post-pulses by more than an order of magnitude, i.e. $\varrho>10$. The highest attosecond pulse intensities are found for $\lambda_1=1300\:$nm and 1400~nm, of which the latter will be the practically most advantageous because it offers a broader CEP-range ($\Delta\CEP=1.7\:$rad) within which an IAP is retained (cp. figures \ref{fig:CEPscansR1}b and c).

The obtained attosecond pulse intensities are two orders of magnitude higher than in the APT with similar photon energy obtained with single-colour drivers (to be generated by OPA) with the same total pulse energy and peak intensity (cp. figure \ref{fig:singlecolours}). This underlines the extremely useful temporal gating action of well-designed two-colour drivers: we do not force the isolation of a very weak attosecond pulse by picking an ever so narrow spectral slice in the very cutoff region, but we obtain, from $\gtrsim10\:$-eV-wide spectra, IAP orders of magnitude more intense than the APT generated with a single-colour driver. Furthermore, reaching higher photon energies, like the $\approx100\:$eV in the examples presented here, with ``efficient'' HHG gas media offering higher recombination cross sections, like argon, generally requires longer driving wavelengths that need to be generated by OPA. It is thus certain that a two-colour field composed of the OPA signal and part of the pump laser emission can be generated with higher total pulse energy. Not only does the two-colour field reach the same high HHG cutoff with higher efficiency, but it will contain more driving energy to begin with, further increasing its advantage in terms of attainable attosecond pulse energy and / or intensity.

For a practical implementation, it will be interesting to know how much the relative weight of the auxiliary colour component can be decreased while still keeping the temporal gating effect intact. While the factorization in equation \ref{eq:factorization} is valid only for equal amplitudes of the beating colour-components, we find that the conclusions drawn from it remain valid guides also for a much weaker auxiliary component. Having repeated the same numerical calculations as shown in figures \ref{fig:CEPscansR1} and \ref{fig:selectedCEPsR1} for $R=1/2$ (results shown in the Supplementary Information), we find that, 
except for the lowest $\lambda_1=1200\:$nm, which no longer leads to an IAP, the results are very similar to those obtained for $R=1$ (attosecond pulse intensities remain the same within $\pm$10\%, the CEP range for IAP generation for $\lambda_1=1400\:$nm remains $\Delta\CEP=1.7\:$rad), if one adapts the position of the spectral filter for the attosecond pulses to the small $\approx5\:$eV decrease of the achieved cutoff photon energies. Upon further lowering the peak intensity ratio to $R=1/9$ (results shown in the Supplementary Information), we find the temporal gating to become significantly less effective, i.e. the CEP-range for which the central driving sub-pulse generates a single attosecond pulse only is now significantly reduced and the emission from the adjacent sub-pulses is less strongly suppressed. Nonetheless, with the spectral filter positions adapted to the further lowered cutoff photon energies, we can still find optimal CEP-values that allow generating an IAP for the auxiliary wavelengths $\lambda_1=1300\:$nm, 1400~nm, 1600~nm  and 1800~nm, with IAP peak intensities reduced by a factor $\lesssim3$ compared to the $R=1$ and $R=1/2$-cases. The auxiliary wavelength $\lambda_1=1400\:$nm is still identified as the best choice, uniting highest attosecond pulse intensity and suppression of satellites. The requirement on CEP-stability ($\Delta\CEP=1\:$rad) is now more stringent than for the cases of higher relative intensity of the auxiliary colour component, but still well achievable in real-world experimental setups \cite{Haessler2014perfectwave}. It will thus be possible to relax the required OPA output energy and retain the majority of the driving energy in the fundamental laser pulse.

\begin{figure}
	\includegraphics[width=.9\textwidth]{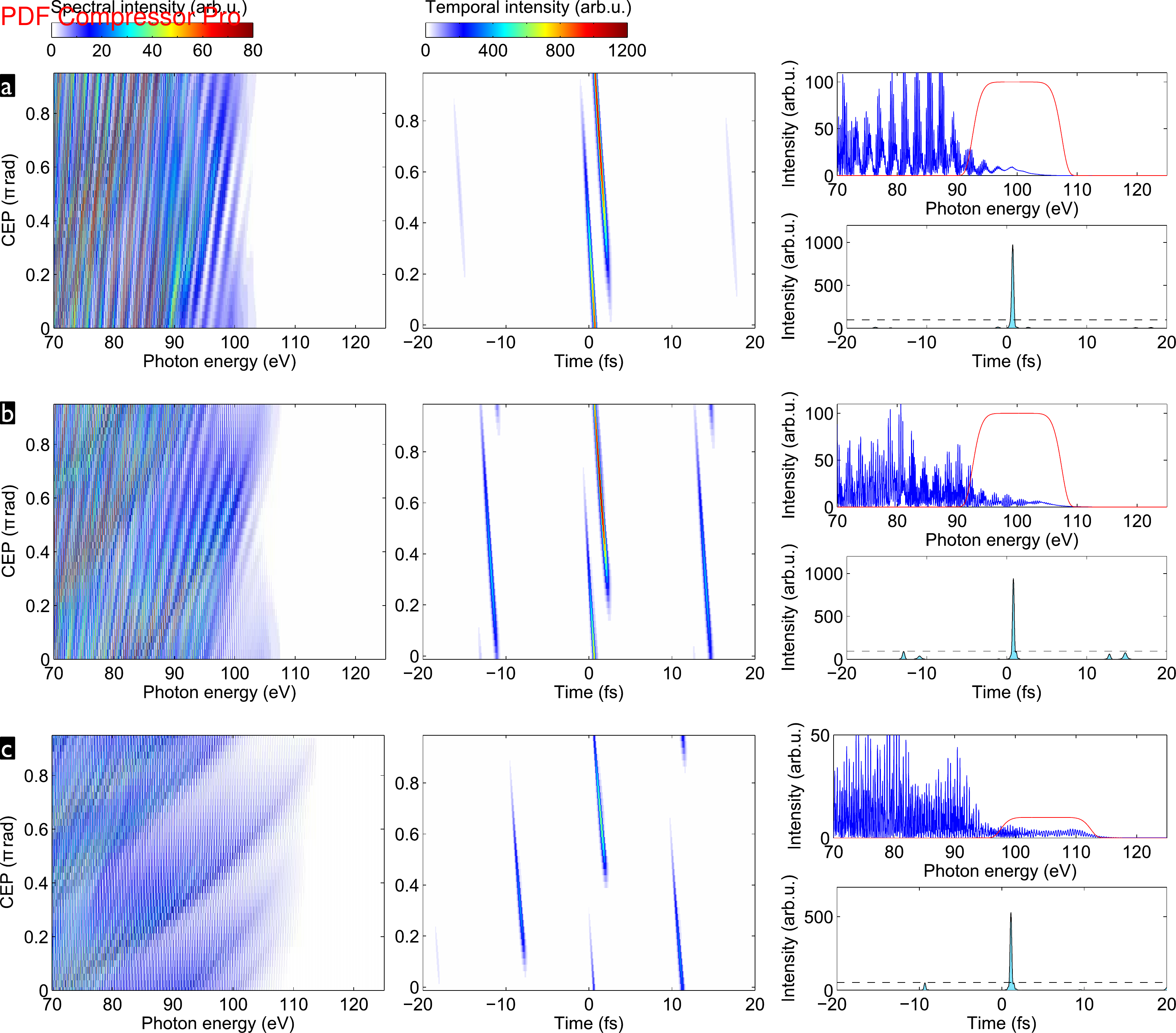}
	\caption{CEP-dependence of the generated high harmonic spectrum (left) and attosecond pulses (middle) for the combination of a 180-fs, 1030-nm fundamental with an auxiliary \emph{non-integer-}$N$ 1300-nm (a), 1400-nm (b), or 1600-nm (c) pulse of  60~fs duration. The pulses have peak intensities of $I_0 = 0.5\times10^{14}\:\mathrm{W\,cm^{-2}}$ and $I_1 = 0.5\times10^{14}\:\mathrm{W\,cm^{-2}}$, respectively, \emph{i.e. $R=1$}. For the attosecond pulses, the cutoff region has been selected by a 15-eV wide normalized super-Gaussian filter centered at 100~eV (a,b) or 105~eV (c), shown by the red line superposed to the spectra on the right. Spectra and attosecond pulses for the selected optimal $\CEP=0\:$rad (a),  $\CEP=0.9\pi\:$rad (b) and $\CEP=0.8\pi\:$rad (c) (right). The dashed line marks the 10\%-level of the intensity of the strongest attosecond pulse.  Outside of the shown time window, the attosecond emission is suppressed to below a $10^{-4}$-fraction of the strongest attosecond pulse's peak intensity.}
	\label{fig:CEPscansR160180}
\end{figure}
Having shown the effectiveness of two-colour temporal gating for 60-fs pulses and having identifed the non-integer-$N$ auxiliary wavelengths $\lambda_1=1300\:$nm, 1400~nm and 1600~nm as most promising choices, we can now explore even longer pulse durations approaching realistically achievable values for Ytterbium-based amplifiers~\cite{Pugzlys2009,Andriuk2011hollow,Ricaud2010YbCaF2regen,amplight,Siebold2008,Hornung2013Polaris}. Since we have found the temporal gating to be very robust against variations of the relative strength of the auxiliary colour component, we can very well consider different pulse envelope durations for the two colour components. This is experimentally relevant since the OPA output can be significnatly shorter than the pump pulses (see section \ref{sec:expimpl}). Figure \ref{fig:CEPscansR160180} shows results for 180-fs fundamental and 60-fs auxiliary pulses (with superposed envelope maxima), with $R=1$ so as to maximise the gating efficiency. This means that the fundamental component now contains three times more energy per pulse, of which however only the central third overlapping the auxiliary pulse is effectively contributing to HHG. As could be expected, we find in comparison to the case where both colour components have 60-fs envelopes (figures \ref{fig:CEPscansR1} and \ref{fig:selectedCEPsR1}) that the action of the central sub-pulse is barely modified while the suppression of the attosecond emission from the adjacent sub-pulses is less efficient. It is however still sufficient for the generation of IAP, albeit within narrowed CEP-ranges: $\Delta\CEP=1\:$rad for $\lambda_1=1300\:$nm, $\Delta\CEP=0.1\:$rad for$\lambda_1=1400\:$nm, and $\Delta\CEP=0.4\:$rad for $\lambda_1=1600\:$nm.

Since a 1300-nm auxiliary wavelenegth is not easily generated in an OPA pumped by a 1030-nm laser (see section \ref{sec:expimpl}), and the CEP-ranges found for the longer auxiliary wavelenegths are incompatible with real-world CEP-jitter, further enhancement of temporal gating will be required for IAP generation with experimentally realistic driver parameters.

\subsubsection{Three-colour gating and enhancement}

Adding a third colour component---the second harmonic of the fundamental--will provide this required enhancement. We extend equation (\ref{eq:twocolourfield}) to
\begin{equation}
		 E(t) = f_0(t) \mathrm E_0 \cos(\omega_0 t + \CEP) + f_0(t) E_2 \cos[2(\omega_0 t + \CEP) + \varphi_2]  + f_1(t) E_1 \cos(\omega_1 t + \CEP),
	\label{eq:threecolourfield}
\end{equation}

i.e. we assume for simplicity that the second harmonic has the same pulse envelope, $f_0(t)$, as the fundamental. The second harmonic has a phase delay $\varphi_2/(2\omega_0)$ with respect to the fundamental wave. Since it is independent of the CEP, it can be considered perfectly stable even in realistic experimental conditions~\cite{Dudovich2006Measuring}.  

\begin{figure}
	\includegraphics[width=.4\textwidth]{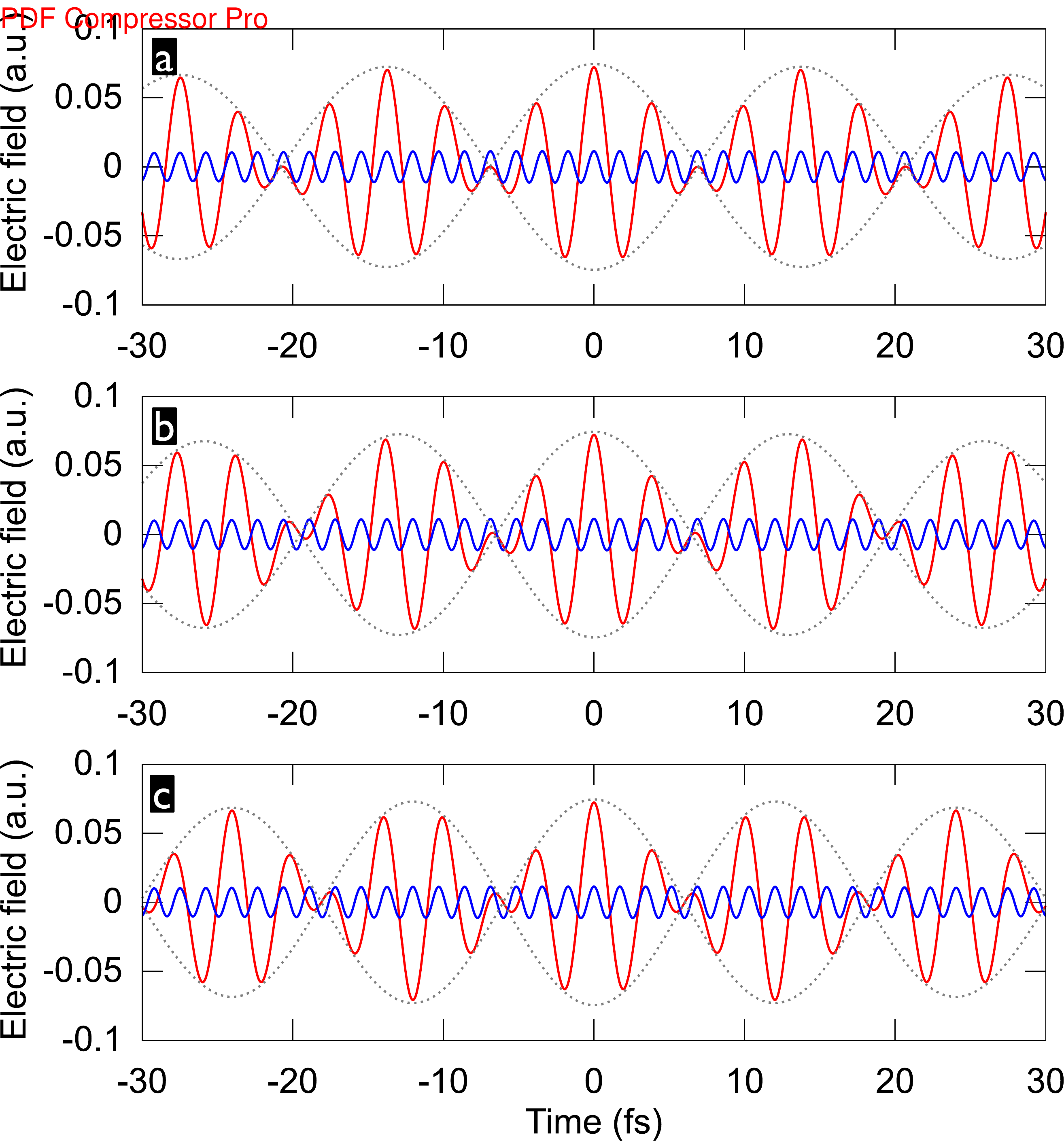}
	\caption{Electric field of two-colour pulses composed from a 1030-nm fundamental and a 1370-nm ($N=7$) (a), 1400-nm ($N=6.6$) (b), or 1440-nm ($N=6$) (c) auxiliary component, with $\CEP=0$ and each a peak intensity of $0.48\times10^{14}\:\mathrm{W\,cm^{-2}}$ ($R=1$) (red line).  As a guide to the eye, the gray dashed lines shows the ``envelope-term'' of the two-colour few-cycle-pulse train. Superposed is a 515-nm component, with $\varphi_2=0$ and peak intensity $0.05\times10^{14}\:\mathrm{W\,cm^{-2}}$  (blue line).}
	\label{fig:add3rdcolour}
\end{figure}
\begin{figure}
	\includegraphics[width=.9\textwidth]{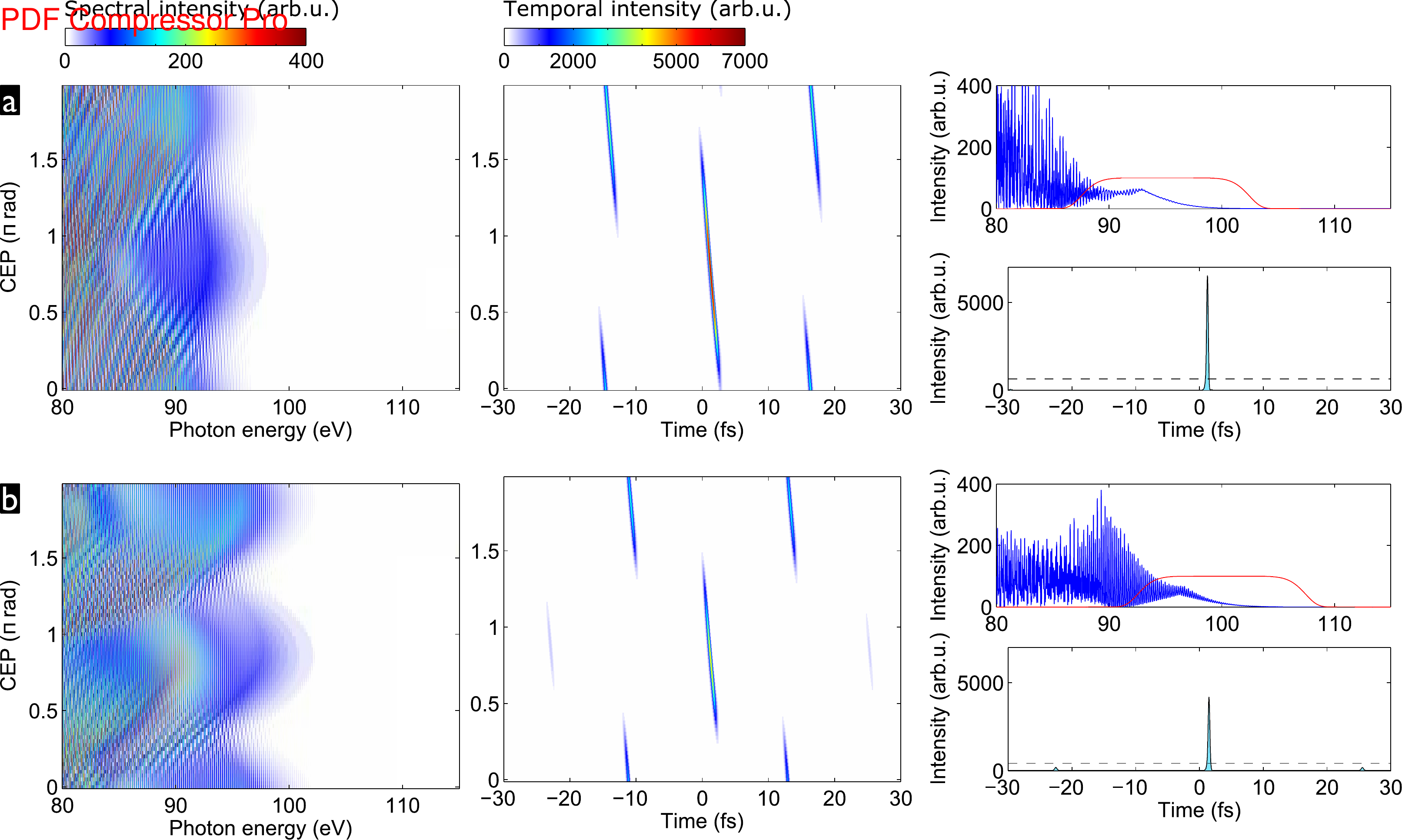}
	\caption{Attosecond pulse generation driven by a three-colour field composed of a 180-fs, 1030-nm fundamental pulse, its 180-fs, 515-nm second harmonic with $\varphi_2=0.1\pi$, and an \mbox{\emph{even}-$N$} auxiliary pulse with 100-fs duration and $\lambda_1=1325\:$nm (a), $\lambda_1=1440\:$nm (b). The pulses have peak intensities of $I_0 = 0.63\times10^{14}\:\mathrm{W\,cm^{-2}}$, $I_2 = 0.06\times10^{14}\:\mathrm{W\,cm^{-2}}$, and $I_1 = 0.31\times10^{14}\:\mathrm{W\,cm^{-2}}$, respectively, \emph{i.e. $R=1/2$}. CEP-dependence of the generated high harmonic spectrum (left) and attosecond pulses (middle). For the attosecond pulses, the cutoff region has been selected by a 15-eV wide normalized super-Gaussian filter centered at 95~eV (a) or 100~eV (b), shown by the red line superposed to the spectra on the right. Spectra and attosecond pulses for the selected optimal $\CEP=0.8\pi\:$rad (a) and  $\CEP=0.7\pi\:$rad (b) (right). The dashed line marks the 10\%-level of the intensity of the strongest attosecond pulse. Outside of the shown time window, the attosecond emission is suppressed to below a $10^{-6}$-fraction of this maximum intensity.}
	\label{fig:CEPscans100180180evenN}
\end{figure}

The enhanced gating with this third colour component is illustrated in figure \ref{fig:add3rdcolour} for different auxiliary wavelengths around the value of 1400~nm, which we had found to provide a good compromise between wide inter-pulse spacing of the few-cycle-pulse train and short few-cycle duration. On the scale of the individual few-cycle pulses, the $2\omega_0$-component functions as a ``detuned second harmonic'' of the carrier wave~\cite{Merdji2007}, effectively ``de-activating'' all but a single half-cycle of the carrier wave for HHG. This doubles the periodicity of the CEP-dependence of HHG from $\pi$ (as in figure \ref{fig:CEPscansR160180} and earlier) to $2\pi$ (see figure \ref{fig:CEPscans100180180evenN}). On the time-scale of the ``envelope-term'' of the few-cycle-pulse train, the effect depends on $N$. For integer $N$ (figures \ref{fig:add3rdcolour}a,c),  the $2\omega_0$-component has the same phase at every few-cycle-pulse maximum. Therefore, for odd $N$, its phase relative to the carrier-wave of the few-cycle-pulse train is also the same at each few-cycle-pulse maximum, whereas it changes by $\pi$ from one few-cycle-pulse maximum to the next for even $N$. This means that if for a given phase $\varphi_2$, for which the $2\omega_0$-component enhances HHG in the central few-cycle pulse, it will have a quenching effect in the adjacent few-cycle-pulses if the auxiliary wavelenegth is chosen such that $N$ is even. This is obviously beneficial for temporal gating since it effectively results in the generation of an IAP only every other few-cycle pulse in the train. 

Ihe following calculations, we will thus consider three-colour drivers with the \emph{even}-$N$ auxiliary wavelengths $\lambda_1=1325\:$nm ($N=8$) or $\lambda_1=1440\:$nm ($N=6$). We set a 180-fs intensity-FWHM duration for the fundamental and its second harmonic, while for the auxiliary colour component, we set a 100-fs duration. The peak intensity of the second harmonic pulse is chosen to be 10\% of that of the fundamental, so that $E_0 = [I_\mathrm{tot}/(R+1.1)]^{1/2}$, $E_1 = [I_\mathrm{tot} R/(R+1.1)]^{1/2}$, and $E_2 = [0.1 I_\mathrm{tot}/(R+1.1)]^{1/2}$. HHG spectra and the corresponding attosecond emission have been calculated over the phase ranges $0\leq\CEP\leq2\pi$ and  $0\leq\varphi_2\leq\pi$. Note that these phase-ranges cover the whole relevant parameter space (for an isotropic generation medium) since the driving waveform for $[\CEP,\varphi_2]$ is just the negative of that for $[\CEP+\pi,\varphi_2+\pi]$.

Figure \ref{fig:CEPscans100180180evenN} shows results of these calculations for an intensity ratio $R=1/2$ and a phase $\varphi_2=0.1\pi$, selected because it yields the highest ``fitness'' of the IAP as defined earlier (i.e. best compromise between attosecond pulse intensity and suppression of satellites). Comparing the CEP-dependences with those for the two-colour drivers (figure \ref{fig:CEPscansR1}c, note the different CEP-ranges), one clearly confirms the gating of attosecond pulse generation to a single carrier half-cycle per few-cycle driver sub-pulse. We also find, as expected, that the CEP range for efficient attosecond pulse generation by the central few-cycle sub-pulse is neatly separated from that where the two adjacent sub-pulses generate efficiently. This leaves a very wide CEP-range for IAP generation of  $\Delta\CEP=2\:$rad for both auxiliary wavelengths. The IAP duration is 320~as in both cases, without considering possible compensation of their intrinsic GDD (``atto-chirp'')~\cite{Rodrigo2005Alu,Bourassin2011control}.  The same calculations for an intensity ratio $R=1$ give the same results, only with $\approx5\:$eV higher HHG cutoff energies. The enhanced three-colour temporal gating thus enables the generation of IAP even with 100-fs auxiliary pulses.

We also note a further increase by a factor $\approx4$ of the obtained attosecond pulse intensities as compared to the two-colour results (figures \ref{fig:selectedCEPsR1}b,c) at the same total peak intensity. This more than compensates for the higher total driver pulse energy of the considered longer pulses, i.e. the effective driver-to-IAP conversion efficiency is enhanced. As compared to the single-colour drivers (cp. figure \ref{fig:singlecolours}), our calculations predict an enhancement of at about three orders of magnitude. It is however important to note that serious prediction about experimental conversion efficiencies would need to involve propagation calculations of the driver and generated XUV fields in the macroscopic ionizing HHG source medium, beyond the scope of this work. 

The choice of an auxiliary wavelength corresponding to an even-valued $N$ is indeed of key importance. Simulations for $\lambda_1=1300\:$nm ($N=8.6$) and $\lambda_1=1370\:$nm ($N=7$) (results shown in the Supplementary Information) result in no IAP being generated (for all $\varphi_2$) since the CEP-ranges for efficient attosecond pulse generation overlap for the central few-cycle driver sub-pulse and those adjacent to it.

\subsubsection{Analysis of quantum paths}

\begin{figure}
	\includegraphics[width=.9\textwidth]{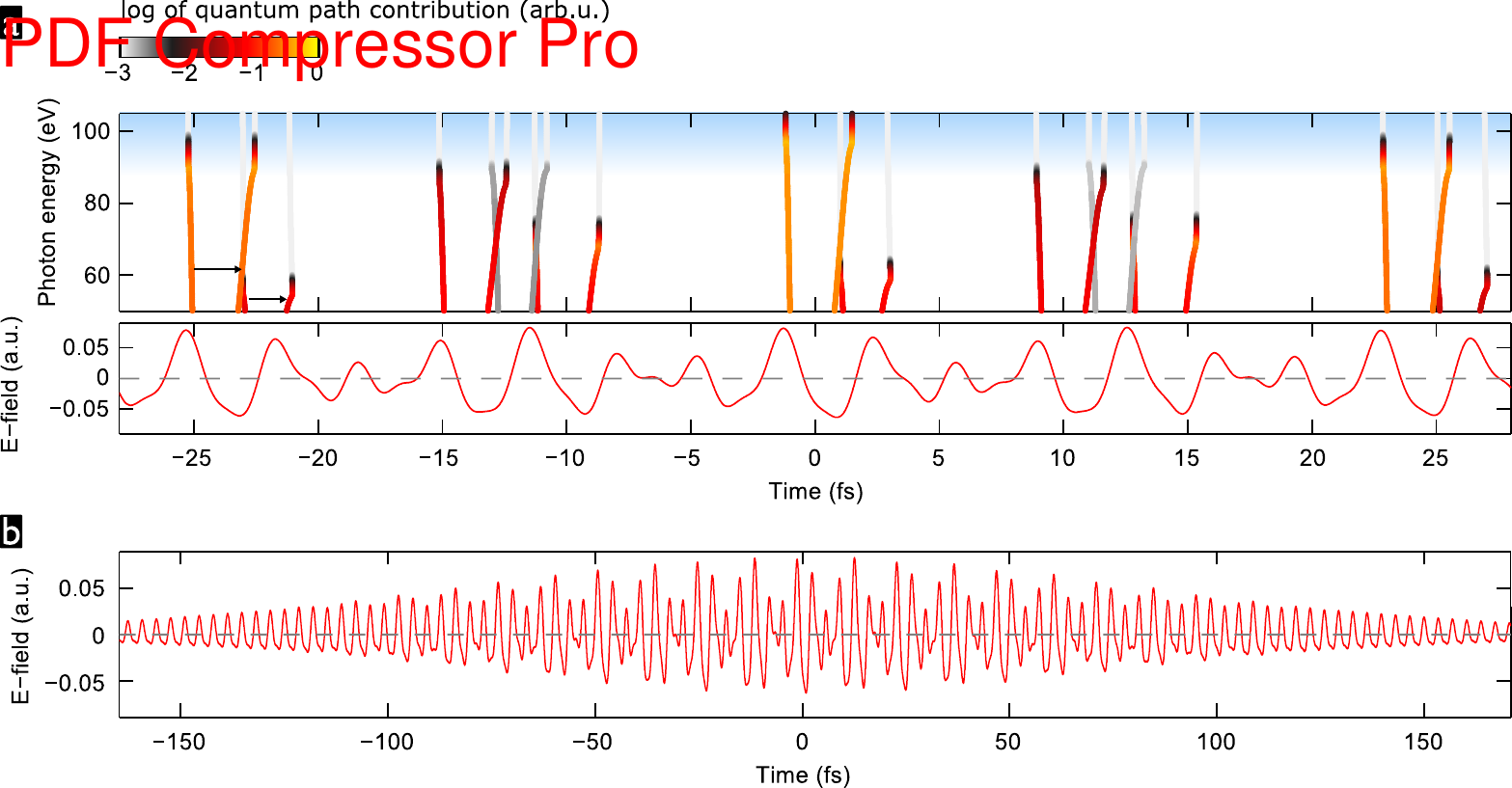}
	\caption{Waveform (zoom-in around pulse peak (a) or full pulse (b)) and driven quantum paths for IAP generation (a). The quantum paths, shown for the short-trajectory branches only, are represented by the real-parts of their ionization and recollision instants. These form matching pairs of curves, connected by arrows for the first of the shown driver sub-pulses. The colour of the data points shows on logarithmic scale the contribution of the corresponding quantum path to the generated HHG intensity. The blue-shaded energy-range marks the spectral fitler used for the selecting an IAP in figure \ref{fig:CEPscans100180180evenN}b. }
	\label{fig:quantumpaths}
\end{figure}
The inner workings of the driver waveforms for optimal temporal gating and efficiency enhancement are well understood by studying the calculated quamtum paths. Figure \ref{fig:quantumpaths} shows the waveform and computed quantum paths correspoding to the results shown in figure \ref{fig:CEPscans100180180evenN}b, i.e. for the \emph{even-}$N$ auxiliary $\lambda_1=1440\:$nm with $\CEP=0.68\pi$ and $\varphi_2=0.1\pi$. The photon energies $>90\:$eV selected for the IAP are generated most efficiently only in a single event during the central driving sub-pulse. This event is launched by the strongests field crest (marked by the ionization instants), which is followed by a strong and broad field-crest efficiently accelerating the returning electron. The sub-pulse is short enough so that the sub-sequent field crest does only lead to much lower recollision energy and thus photon energy. The nearest adjacent sub-pulses have a different, sub-optimal cycle-shape leading to reduced ionization and subsequent electron-acceleration by the driving field. The next adjacent sub-pulses then repeat the cycle-shape of the central one, but the pulse envelopes reduce the peak field strengths so that the $>90$-eV photon energies are only reached with exponentially dropping trajectory contributions in the cutoff region. 

The central sub-pulse has a very similar cycle shape as that described and realized in our recent experimental work~\cite{Haessler2014perfectwave} (where the optimal cycle shape was found to be given by the phases $\CEP=0.85\pi$ and $\varphi_2=0.2\pi$), which demonstrated that such three-colour waveforms can indeed greatly enhance the HHG efficiency by implementing the optimisation-principles laid out by the ``perfect wave'' for HHG~\cite{Chipperfield2009}. While in that work, we have used an \emph{odd}-$N$auxiliary wavelength in order to create perfectly repeating cycle-shapes whose enhancing effect was then studied, here we have targeted the strongest possible temporal gating. The fact that both approaches lead to very similar cycle shapes underlines that not only is the generation of IAP possible with surprisingly long driving laser pulses, but it is at the same time achieved with enhanced efficiency.

\section{\label{sec:expimpl}Experimental feasibility} 

The theory calculations described in this paper were performed for conditions converging to a realistic implementation using currently available laser technology. Ytterbium active laser materials are a particularly promising trend in high average power ultrafast laser development due to low quantum defect and the availability of diode stacks for pumping, which allows extracting unprecedentedly high average power from a laser amplifier~\cite{Klenke2013530W,Haedrich2013nonlinear} and is compatible with scaling towards Joule-level pulse energies~\cite{Siebold2008,Hornung2013Polaris}. The femtosecond pulse duration allows simple implementation of white-light seeded OPA, as in our recent experimental work~\cite{Haessler2014perfectwave}.



\begin{figure}
	\includegraphics[width=12cm]{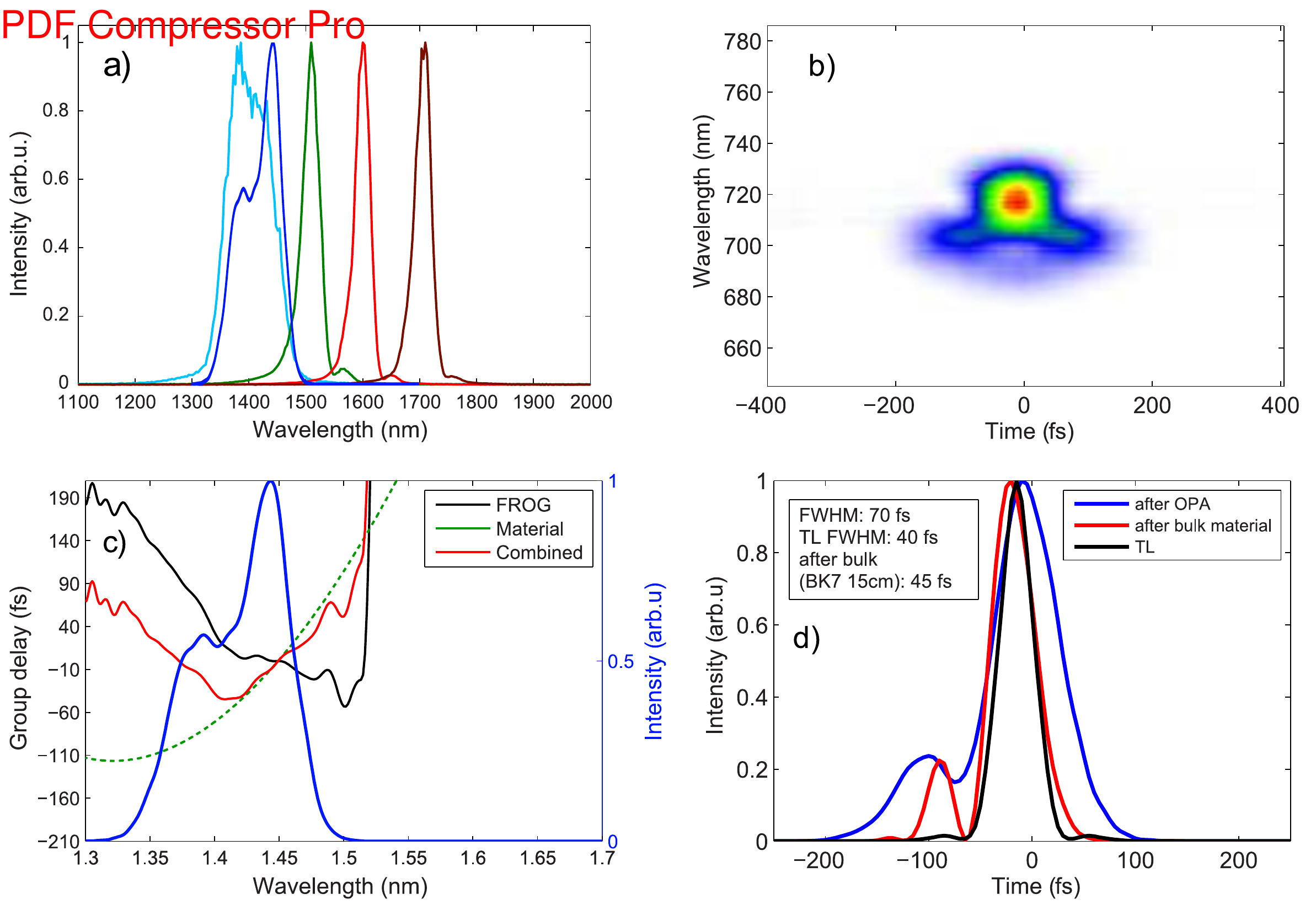}
	\caption{Characterization of the OPA based on KTA and pumped by a 180-fs Yb:CaF2 chirped pulse amplifier system. a) spectra at different central wavelengths,  Panels b)-d) shows the temporal characterization of the pulse from the OPA tuned to 1440-nm central wavelenth and the calculation of the pulse compression using 15 mm thick BK7 glass block. b) measured FROG trace of the pulse directly after the OPA, c) reconstructed group delay, d) temporal pulse profile.\label{fig:OPAcharacterization}}
\end{figure}

As an illustration of such a setup and the OPA tuning range, we show in figure~\ref{fig:OPAcharacterization} spectral and temporal characterization measurements of the signal output from a white-light seeded KTA-based OPA pumped by a 180-fs Yb:CaF$_{2}$ regenerative amplifier. The OPA is built in a two-stage configuration, where the first 4~mm thick crystal is used as a pre-amplifier for the white-light seed generated in a 2-mm thick YAG plate~\cite{Riedle}. The pre-amplified seed (5~$\mu$J) is then subsequently amplified in the second OPA stage with a 2~mm thick crystal. The central wavelength can be tuned within the range of 1400--2000~nm simply by changing the phase matching angle of the KTA crystals. The tuning range on the short wavelength side is limited by the absorption of the idler wave in the KTA crystal above 4000~nm and the related thermal effects. The bandwidth of the signal pulse stays almost the same throughout the tuning range and is limited by the phase-matching bandwidth of the OPA crystal. A notable exception is the $\approx1400$-nm central wavelength where the bandwidth becomes significantly broader (see figure \ref{fig:OPAcharacterization}a) due to matched group velocities of the signal and idler waves~\cite{Cerullo2003RMP}.

The pulses from the OPA are much shorter than the laser pump pulses due to the broad amplification bandwidth and broadband seed pulse derived via white-light generation, as well as the nonlinearity of the parametric gain. In particular for the  amplified signal pulse with 1440~nm central wavelength, for which a frequency-resolved optical gating (FROG) measurement is shown in figure \ref{fig:OPAcharacterization}, we find a 70-fs FWHM duration right after the OPA. This pulse has some residual positive chirp due to self-phase modulation during white-light generation, which can in principle be compensated using a $\approx\:$15-mm BK7 glass block. However, this further increases the third-order phase with the effect that, although the compressed pulse duration (45~fs) is comparable to the transform limit (40~fs), this simple setup comes at the price of relatively poor pulse quality. The high third order phase (apparent as a quadratic group delay in figure~\ref{fig:OPAcharacterization}b) delays the spectral components in the center of the spectrum and only the spectral wings are synchronized. For best compression a more complicated pulse shaping schemescan be used, such as specially designed chirped mirrors or an acousto-optic programmable dispersive filter~\cite{Tournois1997Acoustooptic}.

\begin{figure}
	\includegraphics[width=.9\textwidth]{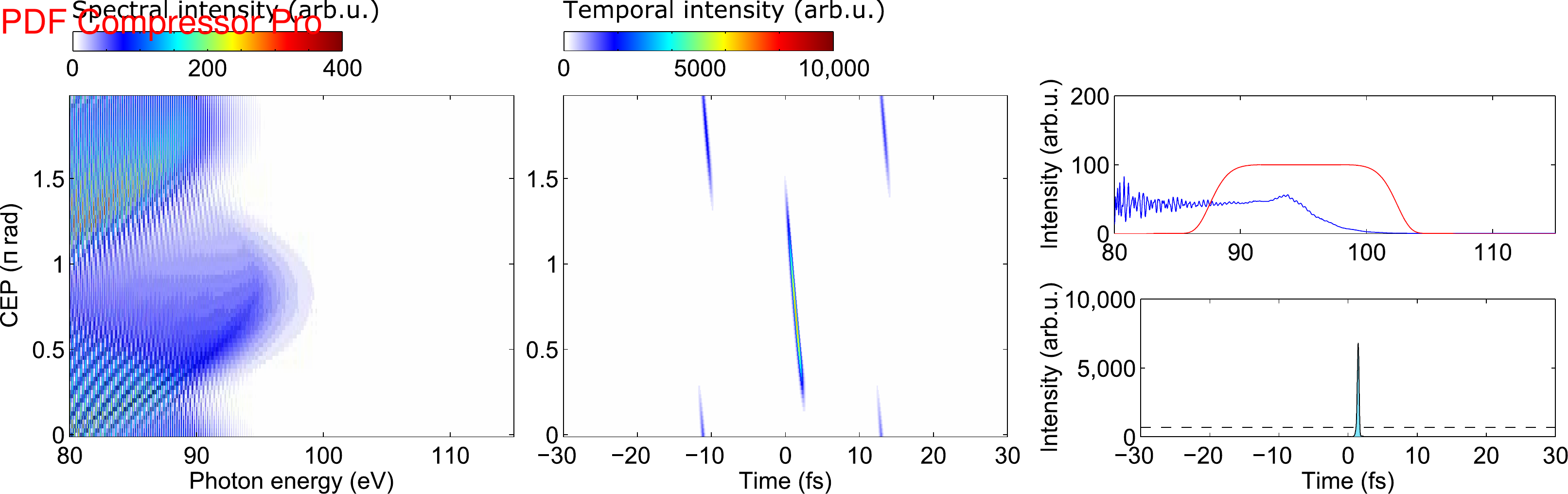}
	\caption{Attosecond pulse generation driven by a three-colour field composed of a 180-fs, 1030-nm fundamental pulse, its 180-fs, 515-nm second harmonic with $\varphi_2=0$, and an \mbox{\emph{even}-$N$} auxiliary pulse with 50-fs duration and $\lambda_1=1440\:$nm. The pulses have peak intensities of $I_0 = 0.63\times10^{14}\:\mathrm{W\,cm^{-2}}$, $I_2 = 0.06\times10^{14}\:\mathrm{W\,cm^{-2}}$, and $I_1 = 0.31\times10^{14}\:\mathrm{W\,cm^{-2}}$, respectively, \emph{i.e. $R=1/2$}. CEP-dependence of the generated high harmonic spectrum (left) and attosecond pulses (middle). For the attosecond pulses, the cutoff region has been selected by a 15-eV wide normalized super-Gaussian filter centered at 95~eV, shown by the red line superposed to the spectra on the right. Spectra and attosecond pulses for the selected optimal $\CEP=0.68\pi\:$rad (right). The dashed line marks the 10\%-level of the intensity of the strongest attosecond pulse. Outside of the shown time window, the attosecond emission is suppressed to below a $10^{-6}$-fraction of this maximum intensity.}
	\label{fig:CEPscans50180180evenN}
\end{figure}

We have thus generated an auxiliary pulse that is much shorter than the 100-fs pulse considered theoretically in figure \ref{fig:CEPscans100180180evenN}. The experimentally realistic situation thus turns out to be quite a bit more comfortable than the extreme case which, according to out single-atom simulation, still supports robust generation of IAP. It sould thus be possible to select a significantly broader XUV-continuum, or to obtain an IAP in a broader CEP-range. This is confirmed by simulations made for a three-colour driver including a 50-fs, 1440-nm auxiliary pulse, shown in figure \ref{fig:CEPscans50180180evenN}, where we find an IAP to be generated within a very broad CEP-range of $\Delta\CEP=\pi\:$rad. 

Our recent experimental work~\cite{Haessler2014perfectwave} has shown that the three-colour drivers we have considered here can in principle be generated with available multi-mJ kHz-repetition-rate Yb-based laser-amplifiers~\cite{Pugzlys2009,Andriuk2011hollow,Ricaud2010YbCaF2regen,amplight} which can be actively CEP-locked~\cite{Balciunas2011cep,Haessler2014perfectwave,Balciunas2014directCEP}. Currently available Yb-based laser technology should thus be able to robustly generate IAP with very high power driver pulses without any need for pulse post-compression. We identify two technical challenges that will be worth tackling for further improvements: 

\emph{(i)} The (long-term) stability of the CEP-locking. While our calculations predict that large jitter around the optimal $\CEP$ is tolerable, the better the experimental stability, the more stable the generated IAP characteristics will be. For a reliable source of IAP, the robust CEP-locking of the Yb-driver-laser to be achieved over long time periods of $\sim$hours. While this has not yet been demonstrated, we do not foresee any principle obstacles. It should be noted that the width of the CEP range for generating an IAP can of course always be increased by shifting the spectral filter towards higher photon energies---though usually at the price of a decreasing IAP peak intensity. In our study, we have avoided to adapt the filter position at each modification of the parameters in order to retain a reasonable comparability of our results.

\emph{(ii)} The lower end of the tunability range of a KTA-based OPA pumped by a 1030-nm laser is at $\lambda_1\approx1400\:$nm. The case of  $\lambda_1=1440\:$nm, considered in figures \ref{fig:CEPscans100180180evenN}b and \ref{fig:CEPscans50180180evenN}, is thus experimentally feasible, while the case of $\lambda_1=1325\:$nm considered in figure \ref{fig:CEPscans100180180evenN}a would require a more complex OPA scheme should be used. For example, a BBO-based OPA pumped by the frequency-doubled 1030-nm fundamental would allow generating a 1325-nm idler wave with very wide bandwidth. The cost is of course a reduced overall OPA efficiency due to the frequency-doubling efficiency typically being below 60\%. A shorter auxiliary wavelength would be however be advantageous since it increases the temporal spacing of the few-cycle sub-pulses and thus improves the suppression of satellite attosecond pulses (the presence of the second harmonic prevents the generation of two attosecond pulses per driver sub-pulse). This would increase the XUV-bandwidth available for the IAP, and thus enable both stronger and shorter IAP (compression of the intrinsic attochirp provided). 

Both mentioned challenges, OPA-tunability and CEP-locking, are essentially removed by a setup such that demonstrated in \cite{Negro2011gating}, where two colour-components with relatively close frequencies are each generated in separate OPA, and derived from a common passively CEP-locked seed. While this arrangement promises to provide great flexibility and more robust CEP-locking, it does of course require a great abundance of fundamental pump laser power and does not make use of the directly available laser output. Since it lifts the requirement of active CEP-locking of the fundamental laser, it can however be implemented with essentially every very-high-power laser system.

\section{Conclusions}\label{sec:concl}

The multi-colour gating method studied here has recently been very succesfully applied in its simpler two-colour variant, albeit without CEP-locking and using an 11-cycle 800-nm fundamental from a Ti:sapphire laser with a 10-cycle 1300-nm auxiliary pulse~\cite{Takahashi2013gigawattatto}. Here we have provided new insight in the mechanism and potential of this method for IAP generation, and show how it can be pushed so far as to enable IAP generation with the extremely interesting and power-scalable Ytterbium-based laser technology, despite the rather long pulse durations. Using currently available CEP-locked Yb-CaF$_2$ amplifiers, this should soon enable the development of novel sources of high-energy IAP with $\sim$kHz repetition rate, making attosecond time-resolved studies feasible for a much broader range of systems.

Our results are also applicable to Ti:sapphire-based HHG drivers. For an 800-nm fundamental, all pulse durations considered in this work simply need to be scaled down by a factor $800/1030 = 0.78$. Also, the achievable photon energies will become smaller due to their $\propto \lambda^2$-scaling: e.g., instead of 100~eV, one would reach 67~eV. The  much shorter pulse durations ($\lesssim30\:$fs, i.e. 11 cycles) directly available from these lasers even in their high-power versions will lead to larger XUV-bandwidths available for the IAP and/or even larger tolerable CEP-jitter.

\section*{Methods}
\subsection*{Non-adiabatic quantum path analysis}

The complex field, $\bfeps(\omega)$, of the high-harmonic radiation with frequency $\omega$ emitted by an atom driven by a laser field is given by (all equations are in atomic units):
\begin{equation}
	\bfeps(\omega) =  \omega^2  \bfx(\omega),
\end{equation}
where $\bfx(\omega)$ is the Fourier-transformed expectation value of the time-dependent atomic dipole moment. Following the fully quantum Lewenstein-model for HHG with the stationary phase (or saddle-point) approximation, developed in refs.~\cite{Lewenstein1994Theory} and \cite{Sansone2004nonadiabatic}, we write:
\begin{multline}
	\bfx(\omega) = \sum_\rms \frac{\rmi 2\pi}{[\mathrm{det}(\partial^2(S + \omega t))]^{1/2}} \left[ \frac{\pi}{\epsilon + \rmi(\tr - \ti)/2}   \right]^{3/2}  
				 \bfE(\ti) \bfd[\bfp_\rms + \bfA(\ti)] \: \bfd^*[\bfp_\rms + \bfA(\tr)]  \\
				  \times \: \exp[\rmi S(\bfp_\rms, \ti, \tr) + \rmi\omega \tr],
\label{eq:dipolesum}
\end{multline}
where the triplets $(\bfp_\rms, \ti, \tr)$ of stationary canonical momentum, ionization instant and recollision instant define the quantum trajectories of the continuum electron over which the sum in Eq.~(\ref{eq:dipolesum}) runs. In this equation, $\bfd [\bfk] = \rmi\, 2^{7/2} (2\Ip)^{5/4} \pi^{-1} \bfk\, (\bfk^2 + 2\Ip)^{-3}$ is the hydrogenic dipole matrix element for bound-free transitions, the star denotes complex conjugation,
\(
	\bfA = - \int_{-\infty}^t \bfE(t'') \rmd t'' 
\)
is the vector potential of the laser field $\bfE(t)$, and the quasi-classical action
\[
	 S(\bfp_\rms, \ti, \tr)  = -\int_\ti^\tr \left[ \frac{[\bfp_\rms + \bfA (t')]^2}{2} + \Ip  \right] \rmd t' ,
\]
with the atomic ionization potential $\Ip$. The determinant
\begin{multline*}
	\mathrm{det}(\partial^2(S + \omega t))] = \left\{   \frac{[\bfp_\rms + \bfA(\tr)] [\bfp_\rms + \bfA(\ti)]}{\tr - \ti} \right\}^2 \\
							- \left\{  -\frac{2(\omega - \Ip)}{\tr - \ti} + \bfE(\tr) [\bfp_\rms + \bfA(\tr)] \right\}  \left\{   \frac{2\Ip}{\tr - \ti} - \bfE(\ti) [\bfp_\rms + \bfA(\ti)] \right\},
\end{multline*}
and $\epsilon$ is a small regularization constant (we set $\epsilon = 10^{-6}$). 

The quantum trajectories are found by solving the system of three saddle point equation resulting from setting $\rmd (S + \omega t) = 0$:
\begin{align}
	\frac{-\bfgreek{\alpha}_{\ti,\tr}}{\tr - \ti} = \bfp_\rms \\
			\ki^2 + 2\Ip = 0, \label{eq:dSdti} \\
			\kr^2 + 2\Ip -2\omega = 0,
\label{eq:dSdtr} 
\end{align}
where $\bfgreek{\alpha}_{\ti,\tr} = \int_\ti^\tr \bfA(t') \rmd t'$. We solve this system of equations numerically (with the \textsc{Matlab} function \textit{fsolve}), using classical solutions for ionization and recollision instants as initial guesses. The solutions $(\bfp_\rms, \ti, \tr)$ are in general complex valued. 

While computing $\bfx(\omega)$ without the stationary phase approximation is perfectly feasible and in fact a widely used standard method for HHG simulations, we have chosen to take the approach via the quantum trajectories since this allows a straightforward connection to classical electron trajectories and thus better physical understanding of the results. Furthermore it lets us pick very easily only the short-trajectory class to take into account for our HHG spectra since this will be the one dominating the (on-axis) emission of a macroscopic HHG medium. The ``quantum path contribution'' which is encoded by the coloured data points in figure~\ref{fig:quantumpaths} is equal to the spectral intensity, $I_\mathrm{HHG}(\omega) = \abs2{\bfeps(\omega)}$, calculated for the respective quantum path only. The full $I_\mathrm{HHG}(\omega)$ results from the interference of quantum paths in the sum (\ref{eq:dipolesum}).

The temporal profile corresponding to a selected spectral range of high-harmonic emission is then found by Fourier transforming $\bfeps(\omega)$. For all temporal profiles shown in this article, we have selected the short-trajectory contribution only since only this will be phase-matched in a macroscopic HHG medium.

%

\acknowledgements
These studies were supported by the ERC (CyFi 280202). S. Haessler acknowledges support by the EU-FP7-IEF MUSCULAR as well as a  ``Chaire Junior''-grant from the ``Laboratoire d'Excellence Physique: Atomes Lumi\`ere Mati\`ere'' (LabEx PALM) overseen by the French National Research Agency (ANR) as part of the ``Investissements d'Avenir'' program (ANR-10-LABX-0039).

%
%

\newpage
\setcounter{figure}{0}   
\renewcommand{\thefigure}{S\arabic{figure}}

\section*{Supplementary information}
\begin{figure}[h]
	\includegraphics[width=0.99\textwidth]{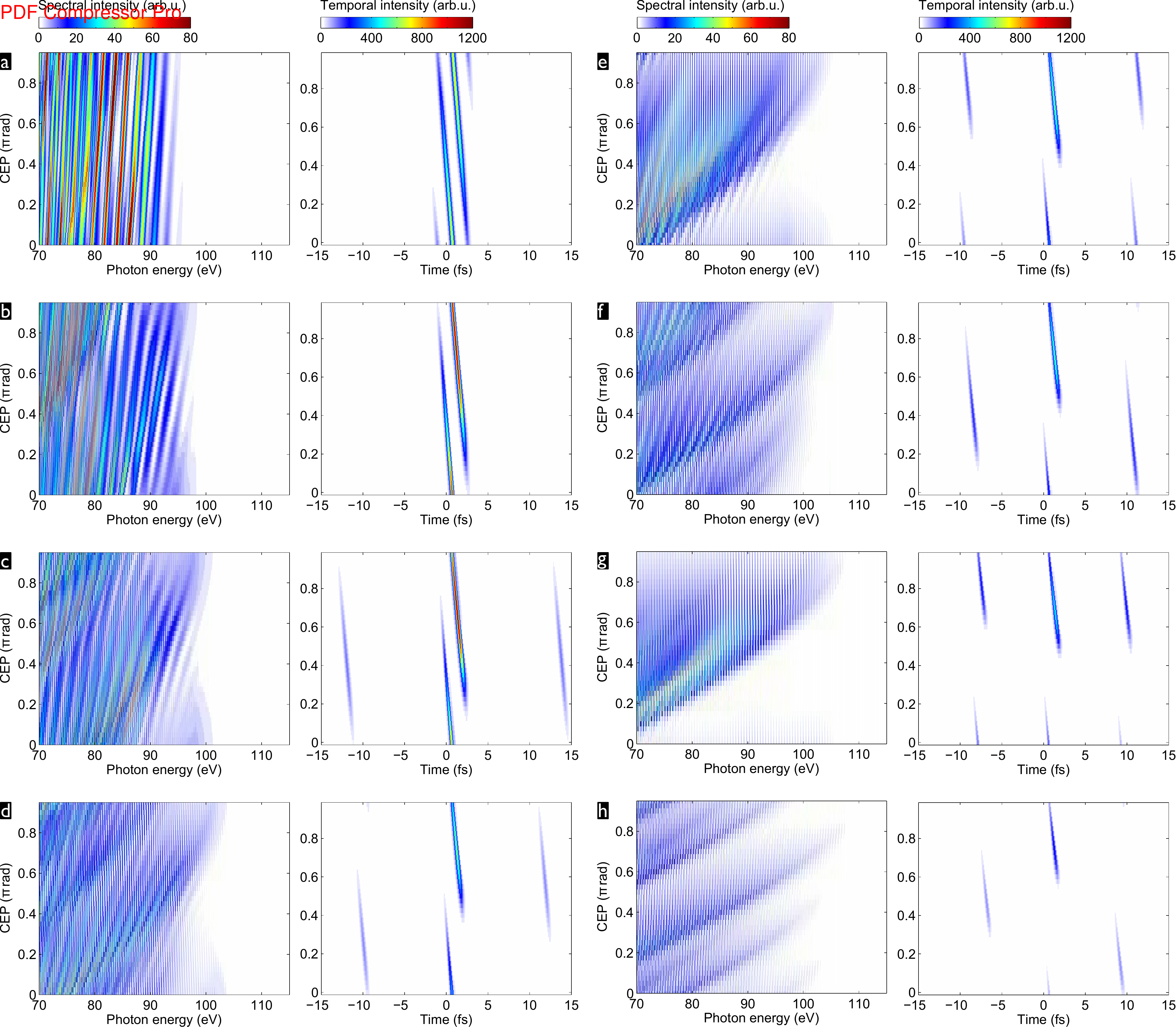}
	\caption{\emph{Equivalent to Fig. 4 of the main paper, but for $R=1/2$.} CEP-dependence of the generated high harmonic spectrum (left) and attosecond pulses (right) for the combination of a 1030-nm fundamental with an auxiliary 1200-nm (a), 1300-nm (b), 1400-nm (c), 1500-nm (d), 1545-nm (e), 1600-nm (f), 1700-nm (g), or 1800-nm (h) pulse, both with a 60-fs duration. The pulses have peak intensities of $I_0 = 2/3\times10^{14}\:\mathrm{W\,cm^{-2}}$ and $I_1 = 1/3\times10^{14}\:\mathrm{W\,cm^{-2}}$, respectively, i.e. $R=1/2$. For the attosecond pulses, the cutoff region has been selected by a 15-eV wide normalized super-Gaussian filter centered at 95~eV (a-c), 100~eV (d-g) or 105~eV (h) (cp. figure \ref{fig:selectedCEPsR05}). Outside of the shown time window, the attosecond emission is suppressed to below a $10^{-3}$-fraction of the strongest attosecond pulse's peak intensity.}
	\label{fig:CEPscansR05}
\end{figure}

\begin{figure}
	\includegraphics[width=0.9\textwidth]{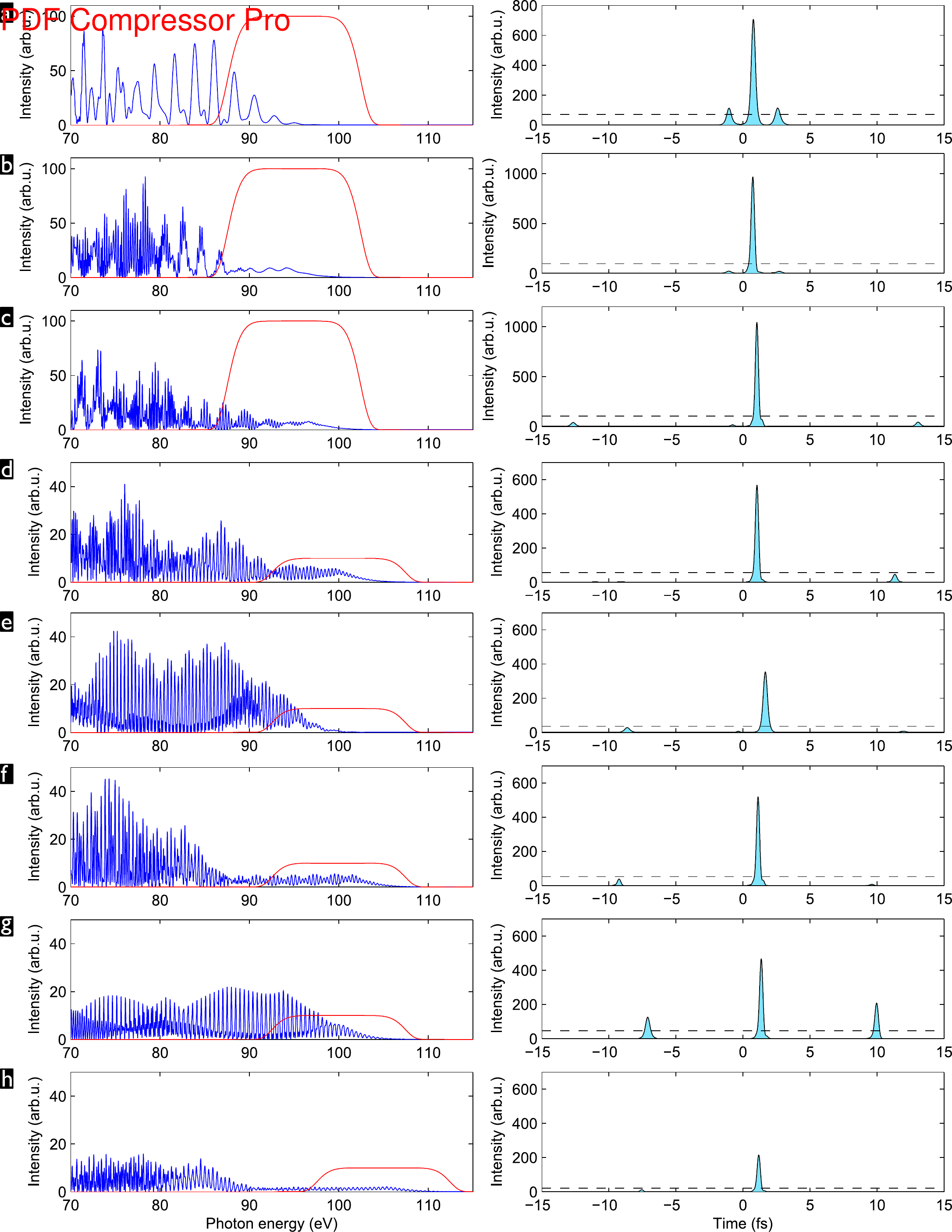}
	\caption{\emph{Equivalent to Fig. 5 of the main paper, but for $R=1/2$.} High harmonic spectrum (left) and attosecond pulses (right) for selected optimal phase delays ($R=1/2$): 
for the auxiliary 1200-nm pulse, $\CEP=0\:$rad (a), for 1300-nm, $\CEP=0\:$rad (b), for 1400-nm, $\CEP=0.8\pi\:$rad (c), for 1500-nm, $\CEP=0.8\pi\:$rad (d), for 1545-nm, $\CEP=0.55\pi\:$rad (e), for 1600-nm, $\CEP=0.8\pi\:$rad (f), for 1700-nm, $\CEP=0.7\pi\:$rad (g), for 1800-nm, $\CEP=0.8\pi\:$rad (h). The red line superposed to the spectra shows the 15-eV wide normalized super-Gaussian filter centered at 95~eV (a-c), 100~eV (d-g) or 105~eV (h), which selects the cutoff region for the attosecond pulses. The dashed line marks the 10\%-level of the intensity of the strongest attosecond pulse.}
	\label{fig:selectedCEPsR05}
\end{figure}

\begin{figure}
	\includegraphics[width=0.99\textwidth]{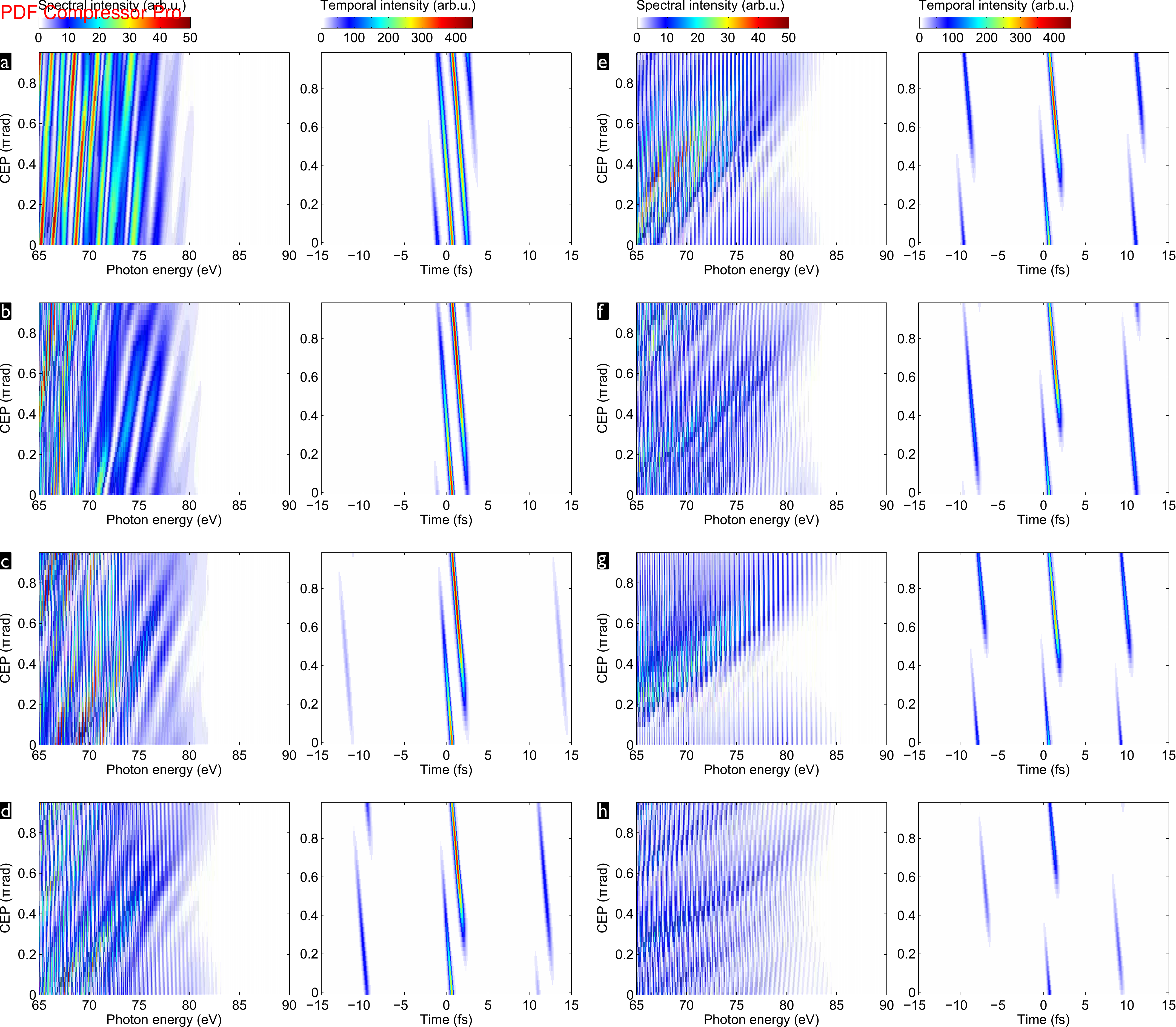}
	\caption{\emph{Equivalent to Fig. 4 of the main paper, but for $R=1/9$.} CEP-dependence of the generated high harmonic spectrum (left) and attosecond pulses (right) for the combination of a 1030-nm fundamental with an auxiliary 1200-nm (a), 1300-nm (b), 1400-nm (c), 1500-nm (d), 1545-nm (e), 1600-nm (f), 1700-nm (g), or 1800-nm (h) pulse, both with a 60-fs duration. The pulses have peak intensities of $I_0 = 0.9\times10^{14}\:\mathrm{W\,cm^{-2}}$ and $I_1 = 0.1\times10^{14}\:\mathrm{W\,cm^{-2}}$, respectively, \emph{i.e. $R=1/9$}. For the attosecond pulses, the cutoff region has been selected by a 15-eV wide normalized super-Gaussian filter centered at 80~eV (a-g) or 85~eV (h) (cp. figure \ref{fig:selectedCEPsR011}).  Outside of the shown time window, the attosecond emission is suppressed to below a $10^{-3}$-fraction of the strongest attosecond pulse's peak intensity.}
	\label{fig:CEPscansR011}
\end{figure}
\begin{figure}
	\includegraphics[width=0.9\textwidth]{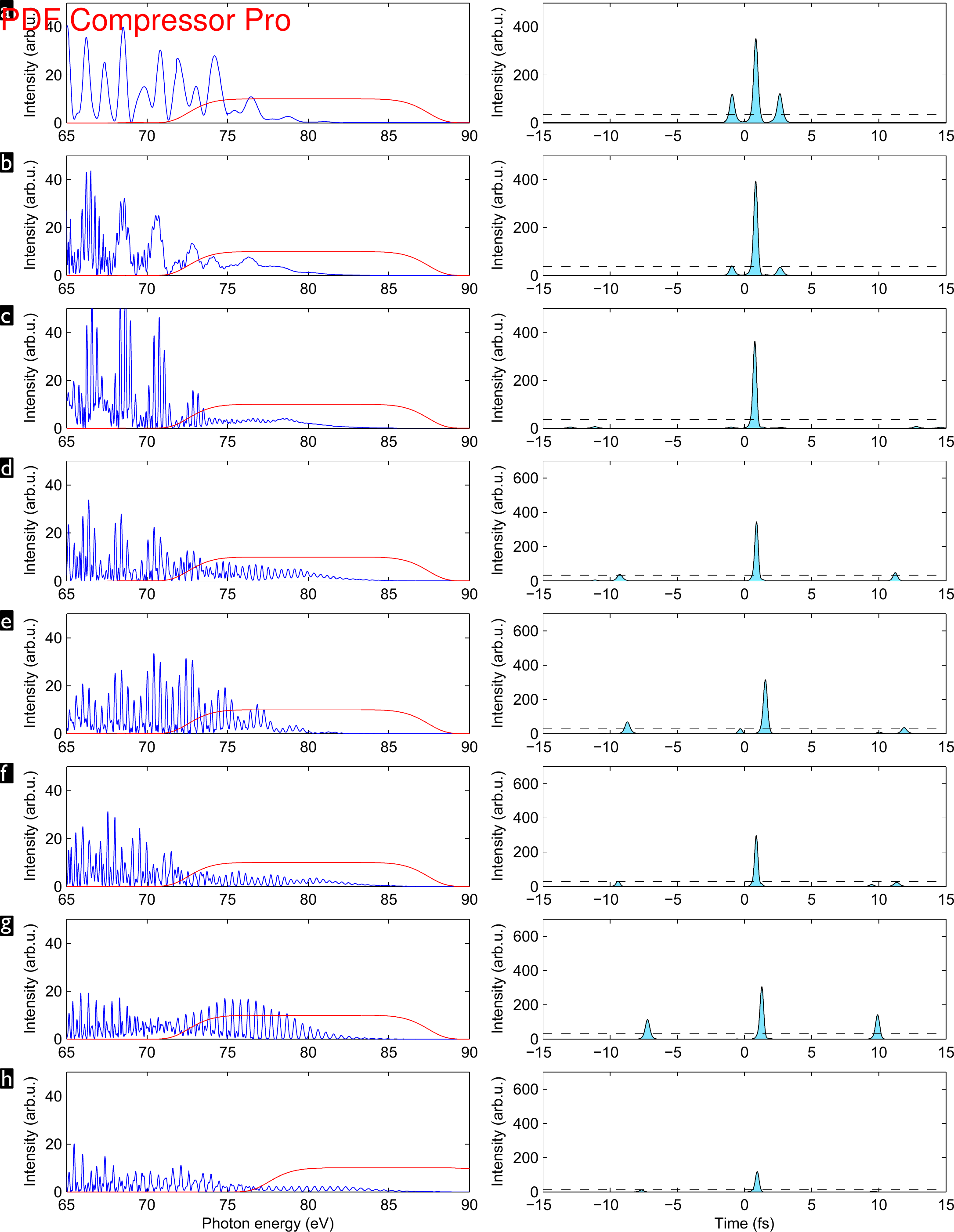}
	\caption{\emph{Equivalent to Fig. 5 of the main paper, but for $R=1/9$.} High harmonic spectrum (left) and attosecond pulses (right) for selected optimal phase delays out of the scans in figure \ref{fig:CEPscansR011} ($R=1/9$): for the auxiliary 1200-nm pulse, $\CEP=0.9\pi\:$rad (a), for 1300-nm, $\CEP=0.9\pi\:$rad (b), for 1400-nm, $\CEP=0.95\pi\:$rad (c), for 1500-nm, $\CEP=0.9\pi\:$rad (d), for 1545-nm, $\CEP=0.55\pi\:$rad (e), for 1600-nm, $\CEP=0.9\pi\:$rad (f), for 1700-nm, $\CEP=0.7\pi\:$rad (g), for 1800-nm, $\CEP=0.9\pi\:$rad (h). The red line superposed to the spectra shows the 15-eV wide normalized  super-Gaussian filter centered at 80~eV (a-g) or 85~eV (h), which selects the cutoff region for the attosecond pulses. The dashed line marks the 10\%-level of the intensity of the strongest attosecond pulse.}
	\label{fig:selectedCEPsR011}
\end{figure}

%

\begin{figure}
	\includegraphics[width=0.75\textwidth]{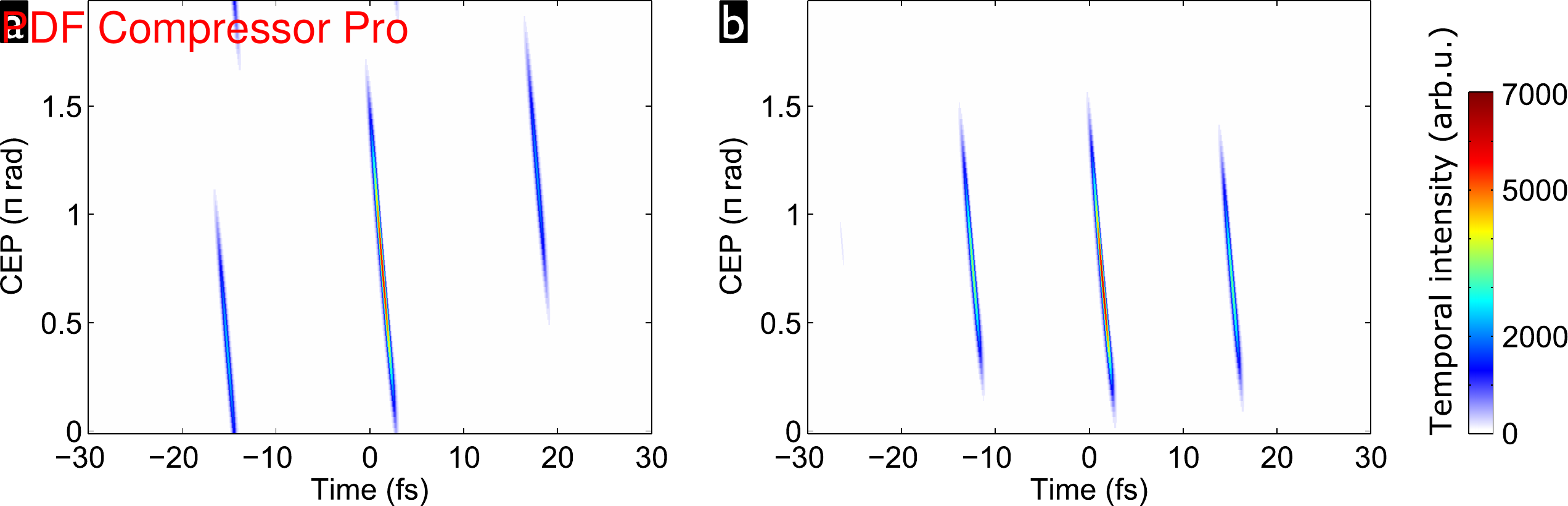}
	\caption{CEP-dependence of the attosecond pulses generated by a three-colour field composed of a 180-fs, 1030-nm fundamental pulse, its 180-fs, 515-nm second harmonic with  $\varphi_2=0.1\pi$ , and an auxiliary pulse with 100-fs duration and $\lambda_1=1300\:$nm (\emph{non-integer} $N=8.6$) (a), $\lambda_1=1370\:$nm (\emph{odd} $N=7$) (b). The pulses have peak intensities of $I_0 = 0.48\times10^{14}\:\mathrm{W\,cm^{-2}}$, $I_2 = 0.05\times10^{14}\:\mathrm{W\,cm^{-2}}$, and $I_1 = 0.48\times10^{14}\:\mathrm{W\,cm^{-2}}$, respectively, \emph{i.e. $R=1$}. For the attosecond pulses, the cutoff region has been selected by a 15-eV wide normalized super-Gaussian filter centered at 100~eV. No IAP can be generated with these auxiliary wavelenegths.}
	\label{fig:quantumpaths}
\end{figure}

\end{document}